\documentclass[journal]{IEEEtran}
\usepackage{amssymb}
\usepackage{amsmath}
\usepackage{mathrsfs}
\usepackage{graphicx}
\usepackage{subfigure}
\usepackage{dblfloatfix}
\usepackage{chngcntr}
\usepackage[justification=centering]{caption}
\usepackage{algorithm}
\usepackage{algpseudocode}
\usepackage{tabularx,booktabs}
\newcolumntype{C}{>{\centering\arraybackslash}X} 
\usepackage{lipsum}
\usepackage{etoolbox}
\usepackage{tikz}
\usetikzlibrary{tikzmark}
\usetikzlibrary{calc}

\begin{document}

\errorcontextlines\maxdimen

\newcommand{\ALGtikzmarkcolor}{black}
\newcommand{\ALGtikzmarkextraindent}{4pt}
\newcommand{\ALGtikzmarkverticaloffsetstart}{-.5ex}
\newcommand{\ALGtikzmarkverticaloffsetend}{-.5ex}
\makeatletter
\newcounter{ALG@tikzmark@tempcnta}

\newcommand\ALG@tikzmark@start{%
    \global\let\ALG@tikzmark@last\ALG@tikzmark@starttext%
    \expandafter\edef\csname ALG@tikzmark@\theALG@nested\endcsname{\theALG@tikzmark@tempcnta}%
    \tikzmark{ALG@tikzmark@start@\csname ALG@tikzmark@\theALG@nested\endcsname}%
    \addtocounter{ALG@tikzmark@tempcnta}{1}%
}

\def\ALG@tikzmark@starttext{start}
\newcommand\ALG@tikzmark@end{%
    \ifx\ALG@tikzmark@last\ALG@tikzmark@starttext
    \else
        \tikzmark{ALG@tikzmark@end@\csname ALG@tikzmark@\theALG@nested\endcsname}%
        \tikz[overlay,remember picture] \draw[\ALGtikzmarkcolor] let \p{S}=($(pic cs:ALG@tikzmark@start@\csname ALG@tikzmark@\theALG@nested\endcsname)+(\ALGtikzmarkextraindent,\ALGtikzmarkverticaloffsetstart)$), \p{E}=($(pic cs:ALG@tikzmark@end@\csname ALG@tikzmark@\theALG@nested\endcsname)+(\ALGtikzmarkextraindent,\ALGtikzmarkverticaloffsetend)$) in (\x{S},\y{S})--(\x{S},\y{E});%
    \fi
    \gdef\ALG@tikzmark@last{end}%
}

\apptocmd{\ALG@beginblock}{\ALG@tikzmark@start}{}{\errmessage{failed to patch}}
\pretocmd{\ALG@endblock}{\ALG@tikzmark@end}{}{\errmessage{failed to patch}}
\makeatother

\title{A New Uncertainty Framework for Stochastic Signal Processing}

\author{Rishabh~Singh,~\IEEEmembership{Student~Member,~IEEE}
        and~Jose~C.~Principe,~\IEEEmembership{Life~Fellow,~IEEE}
\thanks{R. Singh and J. Principe are with the Department
of Electrical and Computer Engineering, University of Florida, Gainesville,
FL, 32601 USA e-mail: rish283@ufl.edu}}
\maketitle

\begin{abstract}
The fields of signal processing and information theory have evolved with the goal of developing formulations to extract intrinsic information from limited amount of data. When one considers the modeling of unpredictably varying processes and complex dynamical signals with a large number of unknowns (such as those encountered in the fields of finance, NLP, communications, etc.), there is a need for algorithms to have increased sensitivity with short spans of data, while maintaining stochastic generalization ability. This naturally calls for an increased focus on localized stochastic representation. So far, most metrics developed for characterizing signals envision data from entropic and probabilistic points of view that lack sensitivity towards quick changes in signal dynamics. We hypothesize that models that work with the intrinsic uncertainties associated with local data induced metric spaces would be significantly more sensitive towards signal characterization. To this end, we develop a new framework for stochastic signal processing that is based on decomposing the local metric space of the signal in a Gaussian Reproducing Kernel Hilbert Space (RKHS). A major advantage of our framework is that we are able to implement this decomposition on a sample-by-sample basis. The key aspects of our framework are the following: (1) We use a data defined metric related to Parzen density estimation for quantifying the local structure of data in the Gaussian RKHS. (2) We use a quantum description of this metric which consequently introduces uncertainty in the structure of the local kernel space. Since the RKHS has been well established and known for providing universal data fitting capabilities, we submit that local quantifiers of the kernel space data projection could significantly improve acquisition of signal information.\end{abstract}


\begin{IEEEkeywords}
Uncertainty, stochastic, decomposition, metric space, dynamical signals, local quantifiers, kernel space, quantum, RKHS.
\end{IEEEkeywords}

\IEEEpeerreviewmaketitle

\section{Introduction}

\subsection{Information Theory: Historical Perspective}
\IEEEPARstart{A}{n} important achievement of Information Theory (IT) was the quantification of statistical uncertainty with the definition of entropy \cite{1}, which remains the leading methodology in use today. This paper focusses on other possible definitions of uncertainty with physical underpinnings, emanating from the duality between particles and waves in quantum theory, which can be applied to functional spaces but do not require the definition of a statistical measure. However, we first review some critical concepts in the field of IT that have led to the applications of entropy for real world problems. The concept of entropy has been at the forefront of IT ever since its inception in the domain of statistics by Shannon \cite{1}. At its origins, Gibbs entropy has been a well-known metric for describing micro-states in a thermodynamical system. Hence, Shannon’s formulation of this concept as a metric associated with behavior of data opened the doors to a completely new realization of information theory based on the uncertainty formulations in physics. Subsequent to its development, Jaynes formulated a method in the context of statistical mechanics to estimate the probability distribution of data as the distribution which maximizes the entropy of the system of data (assumed as micro-states) based on certain macro-state constraints which are typically chosen as some ensemble properties of data \cite{2}. This work has been widely celebrated in the field of density estimation \cite{Terrell}, \cite{hang}. It also created the foundations of physics based interpretation of information theory. Further developments in information theory and stochastic signal processing have been closely intertwined with entropy and several other derived metrics and generalized forms of entropy \cite{3}, \cite{4}. A significant related work is that of Renyi in 1956 where he proposes a generalized formulation of the Shannon’s entropic measure \cite{5}. Renyi’s entropy has the following form:

\begin{equation}
H_{\alpha}(X) = \frac{1}{1-{\alpha}}log\bigg(\sum_{i=1}^{n}{p_i^\alpha\bigg)}
\end{equation}

Here, $X$ is a discrete random variable with $p_i$ denoting the probability of $X$ taking the value $i$.\par

One of the difficulties of applying IT to machine learning is that the data statistics are not usually known a-priori and hence the IT descriptors of entropy, mutual information and divergence must be estimated directly from data \cite{ITL}. Perhaps the simplest of the approaches is to employ statistical methodologies of density estimation \cite{silverman} and plug-in the estimators in the IT formulas. We will review first some alternatives. 
 
\subsection{Maximum Entropy Method of Moments}
An important extension to Jaynes' work is the maximum entropy method of moments which is useful for the estimation of density functions in systems with undersized samples \cite{9}. The main idea here is to construct an approximate density function of a particular system of samples by imposing constraints unique to the system (typically chosen as the moment means) on the various moments associated with data. It consists of a system of two equations from which one can obtain the Lagrangian multipliers (or the weights associated with various data moments) and the probability densities using numerical techniques such as the Newton-Raphson method. Determination of Lagrangian weights quantify the uncertainty associated with the different moments of data. Several application extensions and analysis of this method have been introduced over the past few decades \cite{10,11,12}. This method of maximizing entropy with respect to the various moments of data is simple and elegant with very few assumptions in its formulation. However, the solution of the different Lagrangian multipliers and density values are not guaranteed since the optimization is based on numerical techniques. The possibility of spurious results increase when large number of moments are used. Furthermore, moment correlations are being ignored which leads to more difficulties in solving the optimization problem.\par

\subsection{Point Estimate Methods}
A highly relevant class of methods involved in the formulation and modeling of data statistics are the point estimation methods (PEM) which make it possible to evaluate statistical moments of data using approximate realizations of their density only at specific points in the input space. Introduced by Hahn, Shapiro and Cox \cite{19,20}, PEM methods, at their onset, usually involved Taylor series expansions of random variables about their means. This involves computation of higher order derivatives of functions and can hence become computationally very expensive. One of the popular earlier works in point estimate methods was that of Rosenblueth where he develops a method for estimating densities of correlated variables by assuming the joint distribution of the variables to be concentrated within a set of $2^n$ hyper-quadrants of the space associated with the variables \cite{21,22}. This method becomes impractical for large numbers of quadrants. Li introduced an improved formulation based on Rosenblueth's method that only needs the first two statistical moments to compute PEM with almost the same accuracy as that of Rosenblueth's method \cite{23}. A thorough mathematical analysis and evaluation of various spatial point processes is provided in \cite{24}. A more recent work by Decreusefond and Flint introduces a generalized formulation for point processes with the only constraint being that the processes consist of Papangelou intensities \cite{25}.\par

Although the statistical underpinning of IT is widely accepted, Crutchfield presents a rather unique non-probabilistic perspective of information theory where he regards the information space as a metric space (instead of a probabilistic one) created by information sources \cite{crutch}. He stresses more on the topological structure (geometry) created by information and develops an information theoretic distance measure in such a space that more closely follows the metric properties of distance as compared to other measures (such as mutual information, Kullback information gain, etc.). This serves as an inspiration for us to treat the kernel space projections of data samples as a metric space instead of a probabilistic one in our framework (as we previously have done in \cite{ITL}). 

\subsection{Advances in Uncertainty Quantification Methods}
The field of uncertainty quantification aims to evaluate the effect of intrinsic uncertainties present in the input signal on the model's performance. Earlier methods of uncertainty quantification were based on large number of Monte Carlo simulations of a model to evaluate its variability \cite{r1}. This is obviously a very expensive computational procedure though reliable in terms of accuracy. To mitigate the high computational cost, surrogate models (or meta-models) were introduced as new response surfaces to give fast evaluations of changes in input statistics \cite{r2,r3}. Two such surrogate models that have gained wide popularity are Gaussian process regression \cite{r4} and polynomial chaos expansion \cite{r5,r6}. Both of these algorithms can be implemented in a non-intrusive manner which means that the model being evaluated can be treated as a black box. Wiener first introduced the idea behind polynomial chaos (PC) algorithms in 1938 when he suggested that Hermite polynomials and homogenous chaos had an important influence in integration theory related to Brownian motion \cite{r8}. However, only recently have PC based surrogate models become widely popular \cite{ghanemSpanos}.\par

A polynomial chaos model involves expression of a stochastic process with finite second order moments $Y(x,t,\theta)$ in terms of orthogonal basis functions (given by Wiener-Askey scheme \cite{r11}):

\begin{equation}
Y(x,t,\theta) = \sum_{i=0}^{\infty}{\alpha_i(x,t)\phi_i(\zeta(\theta))}
\end{equation}

Here, $\phi_i$ represents the basis functions (or stochastic modes) and $\zeta$ represents second order random variables of the stochastic process parameterized by $\theta$. $\alpha_i$ represents the PC coefficients or weights of the stochastic modes. The choice of the family of basis functions depends on the distribution of the random variables ($\zeta$) and is given by the Wiener-Askey scheme. Any chosen basis function family, nevertheless, satisfies the following orthogonality conditions:

\begin{equation}
<\phi_r(\zeta), \phi_q(\zeta)> = \int{\phi_r(\zeta)\phi_q(\zeta)P(\zeta) d(\zeta)} = \delta_{rq}||{\phi_r}||^2
\end{equation}

where $r$ and $q$ represent the basis function orders.\par

The main objective in PC expansion is to determine the coefficients ($\alpha$ values in (2)) corresponding to the different modes. This can be done using intrusive or non-intrusive methods. A popular example of an intrusive method is Galerkin projection, which involves solving a system of deterministic equations where the coefficients are considered as the unknowns \cite{r13}. Galerkin projection first involves generalizing (2) in the following manner:

\begin{equation}
Y(x,t,\theta)) = \textit{L}\bigg(x,t;\sum_{r=0}^{K}{\alpha_i(t)\phi_r(\zeta(\theta))}\bigg)
\end{equation}

Projecting (4) into the different polynomial basis functions $\phi_j$, we obtain:

\begin{equation}
\big<Y(x,t,\theta),\phi_q(\zeta(\theta))\big> = \bigg<\textit{L}\bigg(x,t;\sum_{r=0}^K{\alpha_i(t)\phi_r(\zeta(\theta))}\bigg)\bigg>
\end{equation}

This is a system of $K+1$ deterministic equations. These can be solved using integration of the unknowns (the coefficients) with respect to the particular random variable over a suitably sized stochastic space.\par

An example of a non-intrusive method to solve the PC coefficients is linear regression \cite{r15}. Here the coefficients are computed using an overdetermined least squares problem of the following form:

\begin{equation}
\phi(\zeta_k)\alpha(t) = Y(t,\zeta_k)
\end{equation}

A popular class of 'surprise' quantification algorithms over the past few decades have been those related to Bayesian approaches \cite{13,14,15}. Bretthorst, in his work, shows how the entropy method of moments can be formulated in terms of Bayesian probability theory \cite{16}. Through this approach, Lagrangian multipliers are expressed in terms of their marginalized posterior distributions with respect to the number of multipliers and data. Markov Chain Monte Carlo based techniques are then used in obtaining the solutions. While this approach finds solutions that are more optimal than the method of moments, there are significant computational costs involved with Bayesian approaches and Monte Carlo methods that cannot be ignored. A very popular algorithm in this class that gained significant attention over the past decade is that based on a Bayesian definition of surprise by Itti and Baldi \cite{17,18}. The authors formulate the definition of surprise by using the concept of relative entropy. They evaluate the expected Kullback-Leibler (KL) divergence between the prior and posterior distributions of the model after obtaining new data. The formulation is given below:

\begin{equation}
\begin{aligned}
S(D, \mathscr{M}) = KL(P(M|D),P(M)) \\ 
= \int\limits_{\mathscr{M}}P(M|D)log{\frac{P(M|D)}{P(M)}}dM
\end{aligned}
\end{equation}

Here, $M$ is the set of models in a model space given by $\mathscr{M}$. $D$ represents the observed data. We can infer from the formulation that if an observed data leads to a significant change in the posterior distribution of the model, it will subsequently lead to high KL divergence between the prior and posterior thereby causing a high surprise quantity.\par


\subsection{Quantum Modeling of Stock Markets}
The idea of applying quantum theory to data analysis has been pervasive in the field of econophysics [39-45]. The main motivation for doing so comes from treating dynamical signals with a large number of unknown generating functions (such as stock data) as a mixture of different quantum eigenstates associated with its underlying distribution. The earliest such attempts of using physics to describe data can be traced back to 1933 when Frisch attempted to use concepts of classical physics to model finance related dynamics \cite{111}. One of the first crucial attempts in linking quantum physics with finance was made with the quantum field theory interpretation of financial markets where critical concepts from quantum mechanics such as path integrals and differential manifolds were intensively used in the study of the dynamics of financial markets \cite{666, 777}. Recent work in this field involves non-classical oscillator models of the Chinese stock markets \cite{999}, \cite{1000}, \cite{1300} \cite{1400}.

\subsection{Proposed Framework and Contributions}
We propose a framework for characterizing a real valued signal or time series in terms of its intrinsic dynamical moments by decomposing its point-wise stochasticity. Here, stochasticity refers to the uncertainty associated with the local structure of data functionals, instead of a probability measure. In principle, our methodology can also be linked to Gaussian process theory \cite{r4}, except that we do not operate in the data space with its intrinsic probability law. Instead, we prefer to concentrate on the Reproducing Kernel Hilbert Space (RKHS) equivalent functional representation to quantify the local kernel structure of the data.  Hence our framework is more closely associated with point estimate methods. The hallmark of our framework of signal decomposition lies in its ability to perform sample-by-sample decomposition of stochasticity, which is not possible to implement using current techniques. This is done by using a functional called the information potential field, which is motivated from Parzen density estimation of data using Gaussian kernel windows \cite{ITL}. Hence, due the universal approximation property of the Gaussian RKHS, we are able to provide an inclusive local representation of the data, without making assumptions on the type of distribution or stochastic process associated with the data. This enables our framework to exhibit sensitivity towards local signal characteristics while simultaneously providing a generalized representation over all past samples. Since the information potential field is based on the average of the pairwise distances of a point from all available samples, it becomes feasible to treat the associated functional in RKHS as a force field (where the samples are assumed to be identical “information particles”) and hence formulate a physical description of the information potential field based on the time-independent Schrödinger equation \cite{qip}. This quantum description of the information potential field has significant consequence in our framework since it provides a quantification of uncertainty associated with the local RKHS structure. A crucial feature of this quantum description is that the wave-function is defined in the RKHS where the basis functions keep getting updated with every sample. The time-independent Schrödinger equation hence represents the local structure of the RKHS (which formally corresponds to a point-wise estimate of the PDF in a probabilistic interpretation \cite{fuku}) as a combined representation of updated standing waves at any time instance. Since the temporal dependencies of the signal are embedded in the wave-function along all even order moments (see section III), the Schrödinger equation, despite being time-independent, provides an efficient spatiotemporal characterization of a time series. As is evident from the literature of entropy and stochastic decomposition, uncertainties associated with data consist of a mixture of contributions from different moments of data or its stochastic representation \cite{12}, \cite{21}, \cite{r5}. Hence, we utilize this quantum description of the signal’s local RKHS structure to extract higher order intrinsic modes (eigenstates) associated with its uncertainty and their corresponding eigenvalues. We do this by projecting the wave-function into successively higher orders of Hermite polynomial space at every sample. This is similar to formulations associated with polynomial chaos expansion in the field of uncertainty quantification and it also follows the principles of the well-known solution of the quantum harmonic oscillator. The projections, in our case however, are carried out on a sample-by-sample basis. Thereafter, we compute the corresponding information potentials associated with the various extracted modes. The computed information potentials at the different modes summarize the signal uncertainties (in the RKHS) at these modes. The overall framework is depicted in fig. 1 with the associated formulations and reasoning described in detail in sections II and III. A summary of the main steps involved in our sample based framework is as follows: 

\begin{itemize}
\item Quantum description of the local RKHS space using an information potential field created by the selected kernel (and hence a description of uncertainty). We restrict the analysis to the Gaussian kernel in this paper.
\item Extraction of different modes of the wave-function associated with the quantum description of the kernel space using Hermite polynomial embeddings. 
\item Evaluation of the Laplacians over the wave-function modes.
\item Computation of the information potential along each mode using the computed Laplacians.
\end{itemize}

Apart from the ability to operate on a sample-by-sample basis, our framework for uncertainty decomposition of signals presents several advantages over related methods summarized before. Due to the way the Lagrangians are computed, a fundamental limitation of the maximum entropy method of moments is that the moment correlations embedded in the data are ignored. This limits the analysis of the PDF to a very restricted space close to the space of available samples. Our framework, on the other hand, expands the space of analysis much further away from the space of samples by evaluating dynamical moments of the wave-function, which in itself, already contains information related to all even order temporal dependencies of the data (i.e. pairwise differences between the point under consideration and all the other samples that have occurred over time). This also makes our framework advantageous over some of the work associated with probabilistic interpretations of PCA \cite{prob1},\cite{prob2}, where stochasticity is limited to second order statistics due to assumption of Gaussianity. Moreover, we show in section III that the quantum description of the information potential field in our framework involves computation of the Laplacian which takes into account both temporal and inter-modal local dynamics. Furthermore, Hermitian expansions of the wave-function provide a systematic relationship between the successive moments of the wave-function (and hence between the different quantum states of the information potential field). Hence, our framework maximizes information gain and also provides a more stable basis for evaluating as many moments as needed unlike the method of moments where moment expansions depend on the feasibility of the optimization problem for finding the Lagrangians. We show in the formulation of our framework in section III that the eigenvalues associated with the different modes are empirically defined as a result of simply imposing a constraint on the information potential field corresponding to any mode to be always positive. This highly simplifies the eigenvalue determination process and hence provides an advantage to our framework when compared to polynomial chaos based methods where eigenvalue determination process is an optimization problem and hence faces convergence and computational cost issues depending on the type of numerical technique used. Our proposed framework also has several advantages over the stock modeling methods in the field of econometrics summarized before. All of the aforementioned work on quantum modeling are restricted to only specific properties pertaining to the class of stock market data being considered thereby not providing a general framework. In many of these models (\cite{1300}, for instance), authors make specific assumptions on the stochasticity of data (type of motion, drifts, etc) based on the fluctuations of particular stocks. We make no assumptions on the stochastic nature of data. Additionally, unlike some of the stock models (for instance, in \cite{1000}) where authors use time-dependent quantum formulations of the stock data PDF, we use a time-independent Schrödinger equation since our local functional, defining the wave equation, implicitly contains all temporal dependencies of the data. This use of a time-independent formulation greatly simplifies our framework. 

\section{Background Information}
\subsection{Derivation of Information Potential Field (IPF)}
Renyi’s entropy of order alpha ($\alpha \neq 1, \alpha > 0$) for a continuous random variable x is given by:

\begin{equation}
H_\alpha(X) = \frac{1}{1-\alpha}log\int{p(x)^\alpha dx}
\end{equation}

The case of $\alpha = 2$, Renyi's quadratic entropy, is particular important because it leads to an efficient nonparametric family of estimators using the Parzen window method \cite{parz}. In this case, Renyi’s quadratic entropy becomes just the log of the expected value of the probability density function, i.e.

\begin{equation}
H_2(X) = -log\int{p(x)^2 dx} = -logV(X)
\end{equation}

Let us call the argument of the logarithm $V(X)$, the information potential (IP) of the data set, which is a number that is nothing but the mean value of the PDF. Let us assume that we use a Gaussian window for the Parzen density estimation, with bandwidth (or kernel size) $\sigma$. One can readily estimate directly from experimental data ${x_i, i=1, ..., N}$ the information potential,

\begin{equation}
V(X) = \frac{1}{N^2}\sum\limits_{i=1}^{N}\sum\limits_{j=1}^{N}G_{\sigma\sqrt{2}}(x_i - x_j)
\end{equation}

i.e., the IP is a number obtained by the double sum of the Gaussian functions centered at differences of samples with a larger kernel size. There is a physical interpretation of $V(X)$ if we think of the samples as particles in a potential field, hence the name information potential. Let us define a function over the data space $V(x)$ as

\begin{equation}
V(x) = \frac{1}{N}\sum\limits_{i=1}^{N}G(x - x_i)
\end{equation}

which we will call the information potential field (IPF). $V(x)$ is a continuous function obtained by the sum of Gaussian bumps centered on the samples, which is the estimated PDF $\hat{p}(x)$ obtained by the Parzen window, but it can also be interpreted as a potential field similar to gravity, over the space of the samples. In fact, if we attribute unitary mass to all our samples, we can say that the IPF is always positive and regions of space with more samples will have a larger IP, while regions of the space with few samples will have a lower IP. Here, the shape of the Parzen window will determine the “gravity”, instead of the inverse law of physics. The information potential $V(X)$ in Renyi’s entropy is nothing but the total potential field created by the samples in the data set, i.e. $V(X) = \frac{1}{N}\sum\limits_{j=1}^{N}V(x_j)$.

\subsection{Quantum Information Potential field (QIPF)}
The idea of a potential field (aka probability density) over the space of the samples can be readily extended with quantum theoretical concepts \cite{qip}. The Schrödinger stationary (time-independent) equation for a particle in the presence of a potential field can be written as

\begin{equation}
\frac{\hslash^2}{2m}\nabla^2\psi(x) + \psi(x)[E - V_s(x)] = 0
\end{equation}

where $\hslash$ is the Plank’s constant, $m$ the mass of the particle and the wave function $\psi$ determines the spatial probability of the particle with $p(x) = |\psi(x)|^2$. $V_s(x)$ is the potential energy as a function of position, $E$ corresponds to the allowable energy state of the particle and $\psi$ becomes the corresponding eigenvector. For a set of information particles with a Gaussian kernel, the wave-function for one dimensional information particle becomes,

\begin{equation}
\psi(x) = \sqrt{\frac{1}{N}\sum\limits_{i=1}^{N}G_{\sigma}(x-x_i)}
\end{equation}

We can also rescale $V_s(x)$ such that there is a single free parameter $\sigma$ in (12) to yield

\begin{equation}
H\psi(x) = \bigg(-\frac{\sigma^2}{2}\nabla^2 + V_s(x)\bigg)\psi(x) = E\psi(x)
\end{equation}

Solving for $V_s(x)$, we obtain:

\begin{equation}
V_s(x) = E + \frac{\sigma^2/2\nabla^2\psi(x)}{\psi(x)}
\end{equation}

which we call the quantum information potential field (QIPF) denoted by $V_s(x)$. This can also be simplified as:

\begin{equation}
V_s(x) = E - \frac{1}{2} + \frac{1}{2\sigma^2\psi(x)}\sum\limits_{i}(x-x_i)^2e^{-(x-x_i)^2/2\sigma^2}
\end{equation}

To determine the value of $V_s(x)$ uniquely, we require that $minV_s(x) = 0$, which makes $E = -min\frac{\sigma^2/2\nabla^2\psi(x)}{\psi(x)}$ where $0 \leq E \leq 1/2$. Note that $\psi(x)$ is the eigenfunction of $H$ and $E$ is the lowest eigenvalue of the operator, which corresponds to the ground state. Given the data set, we expect $V_s(x)$ to increase quadratically outside the data region and to exhibit local minima associated with the locations of highest sample density (clusters). This can be interpreted as clustering since the potential function attracts the data distribution function $\psi(x)$ to its minima, while the Laplacian drives it away, producing a complicated potential function in the space. We should remark that, in this framework, $E$ sets the scale at which the minima are observed. This derivation can be easily extended to multidimensional data.\par
We can see that $V_s(x)$ is also a potential function that differs from $V(x)$ in (11) because it is associated with a quantum description of the IPF. The two fields are similar to each other for Gaussian kernels since the derivative of the Gaussian is a Gaussian, but it presents a big advantage because now $V_s(x)$ can be operated independently as a counterpart of the samples, describing the duality between particles (samples) and waves (functionals).

\setcounter{figure}{0}
\begin{figure*}[!t]
\centering
\includegraphics[height = 7cm, width = 14cm]{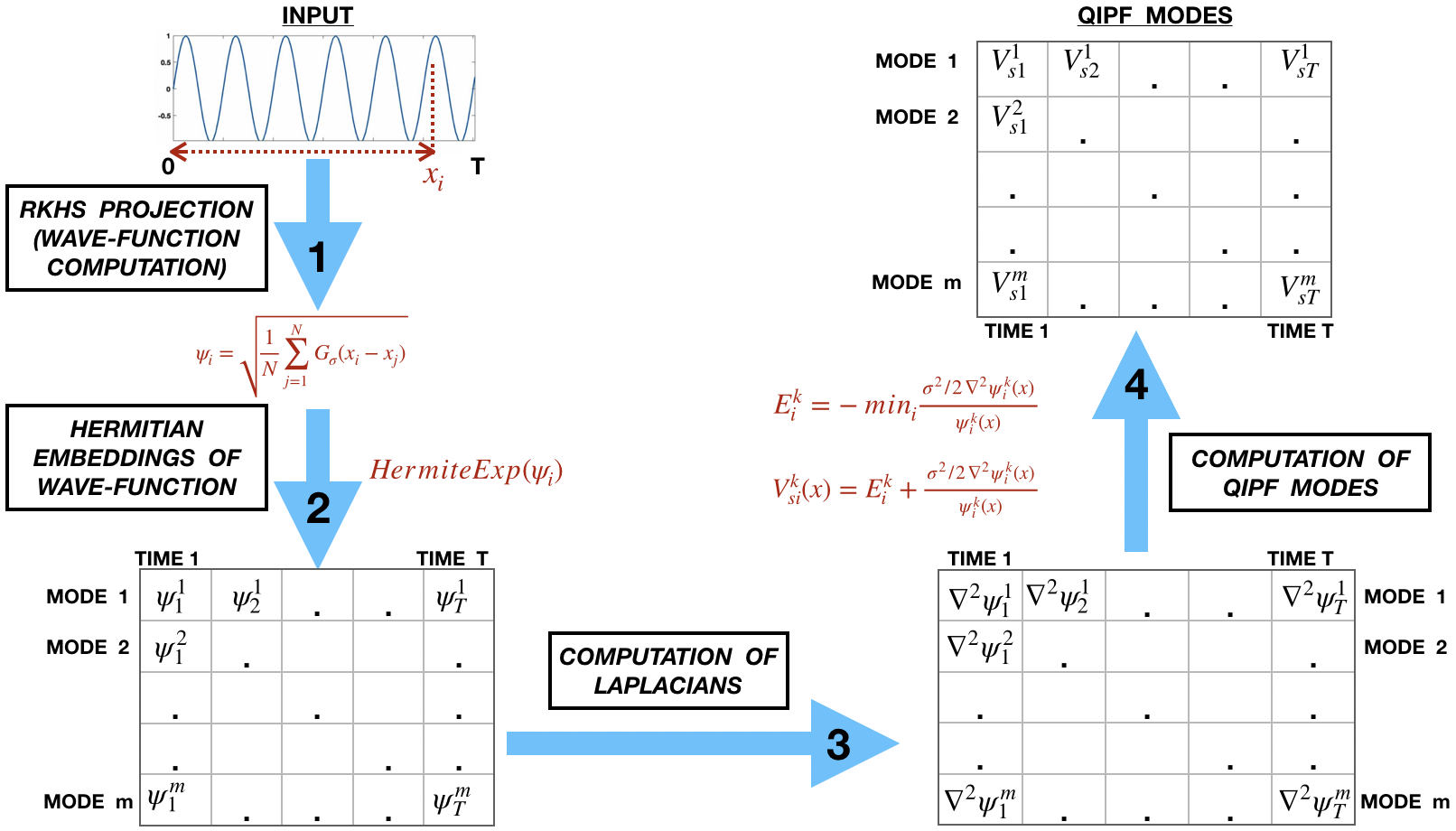}
\caption{Proposed framework for quantum decomposition of information potential field}
\label{fig:1}
\end{figure*}

\section{Proposed Approach: Extracting Modes of Uncertainty in the RKHS}
As one can notice, when we derive information potential as a special case of Parzen's density estimator, it becomes a localized PDF estimator. However, from a deterministic sense, information potential field (IPF) is nothing but a local functional based on the averaged pairwise distance of a point from all the other points in the Gaussian kernel space. From this perspective, its quantum description introduces uncertainty in the functional space associated with the sample location being considered at any point of time: One can consider that at an information particle location, we have no uncertainty (the field collapses to a delta function), but at any point in the RKHS that has no sample, there is an uncertainty that is quantified by the wave-function. Our framework attempts to utilize these concepts to extract the different eigenstates of the QIPF associated with the signal. The main motivation for doing so comes from assuming the signal was created by a quantum physical system that is governed by a large number of unknown forces. Consequently, the extracted eigenstates associated with the quantum framework of the signal would provide a characterization of the intrinsic governing forces acting on the system that produced the signal.\par

As is well known in the field of quantum physics, the time-independent Schrödinger equation describing the dynamics of the quantum harmonic oscillator can be formulated as:

\begin{equation}
\frac{d^2\psi}{dx^2} + \bigg(\frac{2mE}{\hslash^2} - \frac{m^2w^2}{\hslash}x^2 \bigg)\psi = 0
\end{equation}

where the different terms have their usual meaning.
It is widely known that (17) can be solved using the power series method by introducing a dimensionless variable as an expression of $x$ given by $y = \sqrt{\frac{mw}{\hslash}}x$. This yields wave-functions of the different modes, $\psi_n(x)$, that are in fact, consecutive projections of $y$ in the space of Hermite polynomials. These solutions are given as: 

\begin{equation}
\begin{aligned}
E_0 = \frac{{\hslash}w}{2}, &&& {\psi}_0 = {\alpha}_0e^{\frac{-y^2}{2}}\\
E_1 = \frac{3{\hslash}w}{2}, &&& {\psi}_1 = {\alpha}_0(2y)e^{\frac{-y^2}{2}}\\
E_2 = \frac{5{\hslash}w}{2}, &&& {\psi}_2 = {\alpha}_0(4y^2 - 2)e^{\frac{-y^2}{2}}\\
.\\
.
\end{aligned}
\end{equation}

Hence, the solution to the Schrödinger equation for the harmonic oscillator yields infinite eigenfunctions (denoted by $\psi_0, \psi_1, \psi_2...$) that are embedded in the orthogonal space of Hermite polynomials. Their corresponding eigenvalues are denoted by $E_0, E_1, E_2..$.\par

One can obtain similar Hermitian embeddings by projecting the wave-function of the QIPF (13) into the orthogonal spaces of consecutive Hermite polynomials. Our conjecture is that by doing so, we are obtaining the approximate intrinsic modes associated with the local RKHS structure (or the local PDF, from a probabilistic perspective) of the signal as defined by the corresponding information potential field. Hence, by exploiting the Hermitian relationship between the different higher order modes of the QIPF, it becomes possible to construct its eigenfunctions. This is conceptually very similar to working with the moment expansions of the characteristic function in statistics, but without imposing a preset family of basis functions (the complex exponentials). Here, one uses a data centric decomposition instead. The generating function of the Hermite polynomial family is given by:


\begin{equation}
H_n(x) = (-1)^ne^{x^2}\frac{d^n}{dx^n}e^{-x^2}
\end{equation}

Upon inserting the QIPF wave-function in (19), we obtain the following generating function for the wave-functions at different eigenstates:

\begin{equation}
H_n(\psi(x)) = (-1)^ne^{\psi(x)^2}\frac{d^n}{dx^n}e^{-\psi(x)^2}
\end{equation}

The recurrence property of Hermite polynomials allows us to analytically compute successive orders of eigenfunctions without having to use the generating function given by (19). This saves significant computational time when evaluating higher order eigenfunctions of the signal's QIPF in real time by computing the next order of polynomial expansion from current and previous orders using the following relation:

\begin{equation}
H_{n+1}(y) = 2yH_n(y) - 2nH_{n-1}(y)
\end{equation}

Since the Hermite polynomial expansion is being done in the Gaussian kernel space (which is an even function), we only consider even order Hermite expansions \cite{herm}. Thus, on expanding (20) for different even values of n (eigenstates), we get the following series of eigenfunctions in the QIPF:

\begin{equation}
\begin{aligned}
&{\psi}^0 = 1\\
&{\psi}^2 = 4\psi(x)^2 - 2\\
&{\psi}^4 = 16\psi(x)^4 - 48\psi(x)^2 +12\\
&{\psi}^6 = 64\psi(x)^6 - 480\psi(x)^4 +720\psi(x)^2 - 120\\
&.\\
&.
\end{aligned}
\end{equation}

The family of functions thus obtained in (22) form a new space of orthogonal basis functions \textit{within the RKHS totally defined by the time series}. While the Gaussian kernel framework provides us with a with a universal tool for signal representation, the subsequent Hermitian projections (quantified as wave-functions) decomposes the local RKHS structure of the signal in terms of its various underlying moments thereby quantifying uncertainties along those moments.\par

We can normalize the Hermitian expansions such that they satisfy the following relation:

\begin{equation}
\int\limits_{-\infty}^{\infty} e^{-y^2}[ H(y)]^{2}dy = 1
\end{equation}

We now evaluate the QIPF of samples at the different eigenfunction projections of the wave-function using (15) which is generalized here for all orders of expansion:

\begin{equation}
V_{si}^k(x) = E_i^k + \frac{\sigma^2/2 \nabla^{2}\psi_i^k(x)}{\psi_i^k(x)}
\end{equation}

Here, $V_{si}^k$ represents the $k^{th}$ mode of the Schrödinger information potential at sample $i$, $E_i^k$ is the corresponding eigenvalue of the $k^{th}$ mode at sample $i$ and $\psi_i^k(x)$ is the $k^{th}$ mode wave-function value at sample $i$. The Laplacian operator used in the second term in (24) provides a critical advantage to our framework since it efficiently characterizes the local dynamics of the signal along time as well as across the extracted modes. The energy values $E_i^k$ at the different eigenfunction evaluations of information potential are determined empirically by asserting the requirement that $minV_{si}^k(x) = 0$, leading to:

\begin{algorithm}[!b]
\caption{Quantum decomposition of IPF}\label{euclid}
\begin{algorithmic}
\State \textbf{Input:}
\State $x$: Signal
\State $\sigma$: Kernel width
\State $m$: Number of quantum modes
\State \textbf{Initialization:}
\State $\psi$: Wave-function
\State $\psi^2, \psi^4, ... , \psi^m$: Wave-function Hermitian embeddings
\State $V_s^2, V_s^4, ... , V_s^m$: QIPF modes
\State $E^2, E^4, ... , E^m$: Eigenvalue of each mode
\State \textbf{Computations:}
\For {$i = $ 1 to length($x$)}
\State $\psi=0$
\For {$j = $ 1 to i}
\State $\psi \gets \psi + e^{-\frac{(x_i - x_j)^2}{2\sigma^2}}$
\EndFor
\State $\psi_i \gets \sqrt{mean(\psi)}$
\State 
\State $[\psi_i^2, \psi_i^4, ... , \psi_i^m] \gets Hermite  Projections(\psi_i)$
\State $[{\nabla^2}\psi_i^2, ... , {\nabla^2}\psi_i^m] \gets Laplacians$ 
\State 
\For {each mode $k$}
\State $E_i^k = -\min_{q=1...i}\frac{\sigma^2/2{\nabla^2}\psi_q^k}{\psi_q^k}$
\State
\State $V_{s(i)}^k = E_i^k + \frac{\sigma^2/2{\nabla^2}\psi^k}{\psi^k}$
\EndFor
\EndFor
\end{algorithmic}
\end{algorithm}

\begin{equation}
E_i^k = -min_i\frac{\sigma^2/2\nabla^2\psi_i^k(x)}{\psi_n^k(x)}
\end{equation}

$E_i^k$ term gives the energy magnitudes (eigenvalues) of different modes at every sample. This term represents a global quantity in the QIPF since it is a result of constraining the QIPF at the particular mode to be always positive throughout the past samples. An appealing aspect of this term is that it is empirically defined without the use of optimization or regression methods as is done in the eigenvalue determination in polynomial chaos method. Therefore, (24) provides a unified representation of the QIPF at different energy levels or modes. Due to the simplicity of the constraint, it may be argued that the eigenvalues in our case do not provide exact measures of the contribution of each mode. However, we show in our experimental evaluations in section IV that they are capable of uniquely characterizing the dynamics of different signals. The proposed framework is summarized in fig. 1 and Algorithm 1.\par

\section{Simulation Results and Analysis}
In order to study the trends of the extracted QIPF modes in the sample space and analyze their response to various kernel widths, we first perform a spatial analysis of the different QIPF modes. We implement our framework on 500 samples of Lorenz series (which is a chaotic dynamical signal) to extract the first 6 even order modes of the QIPF. All simulations are conducted using MATLAB R2018b on a 2.3 Ghz intel i5 processor machine. The generating functions of the Lorenz series consist of the following mutually coupled differential equations (with $\sigma$, $\rho$ and $\beta$ as system parameters):

\begin{equation}
\fontsize{8pt}{12pt}\selectfont
\begin{aligned}
\frac{dx}{dy} = \sigma(y-x) \\
\frac{dy}{dt} = x(\rho - z) - y \\
\frac{dz}{dt} = xy - \beta{z} 
\end{aligned}
\end{equation}

\setcounter{figure}{1}
\begin{figure*}[!t]
    \centering
    \subfigure[Kernel width = 0.1]{\includegraphics[height = 2.4cm, width=0.32\textwidth]{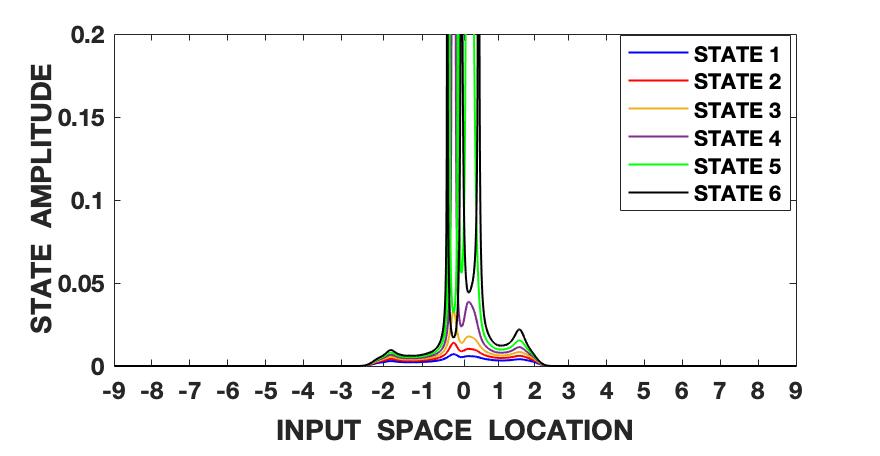}}
    \subfigure[Kernel width = 0.3]{\includegraphics[height = 2.4cm, width=0.32\textwidth]{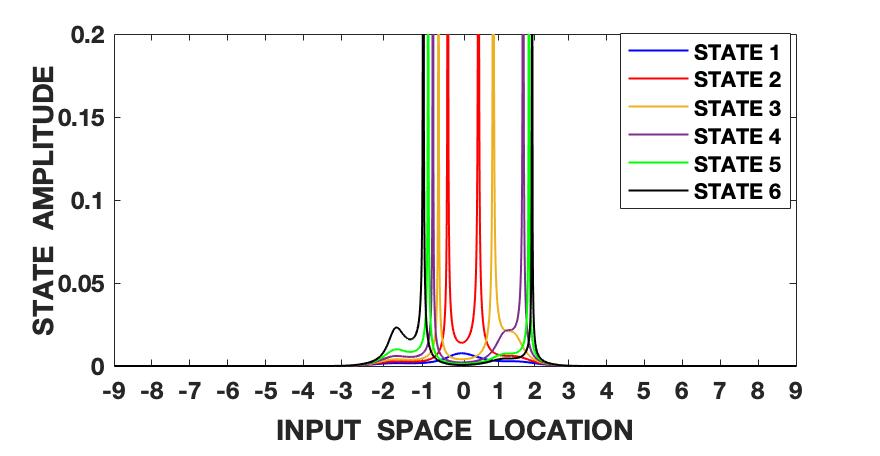}}
    \subfigure[Kernel width = 0.5]{\includegraphics[height = 2.4cm, width=0.32\textwidth]{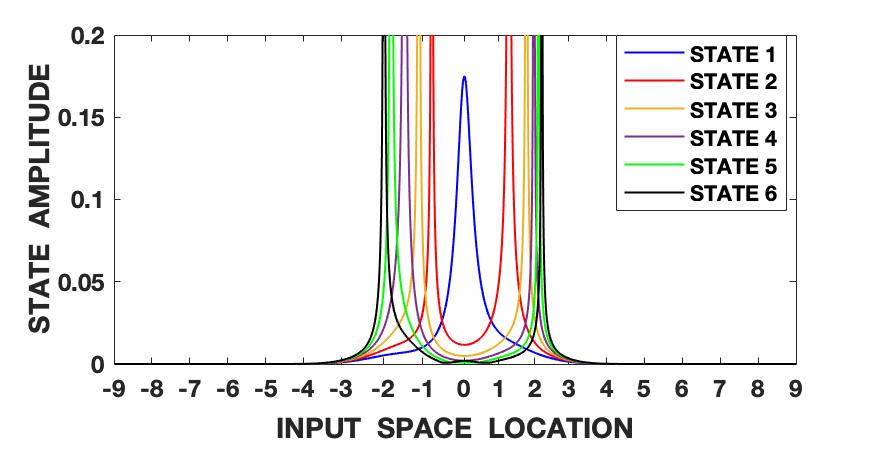}}
    \subfigure[Kernel width = 1]{\includegraphics[height = 2.4cm, width=0.32\textwidth]{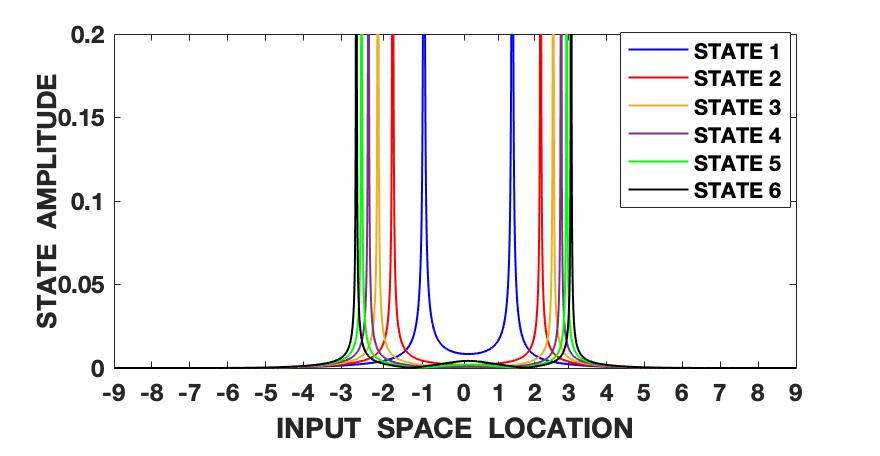}}
    \subfigure[Kernel width = 1.5]{\includegraphics[height = 2.4cm, width=0.32\textwidth]{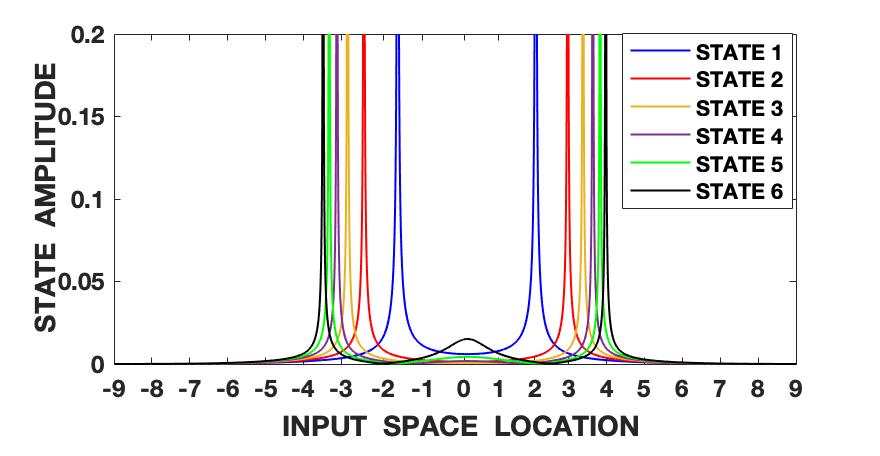}}
    \subfigure[Kernel width = 3]{\includegraphics[height = 2.4cm, width=0.32\textwidth]{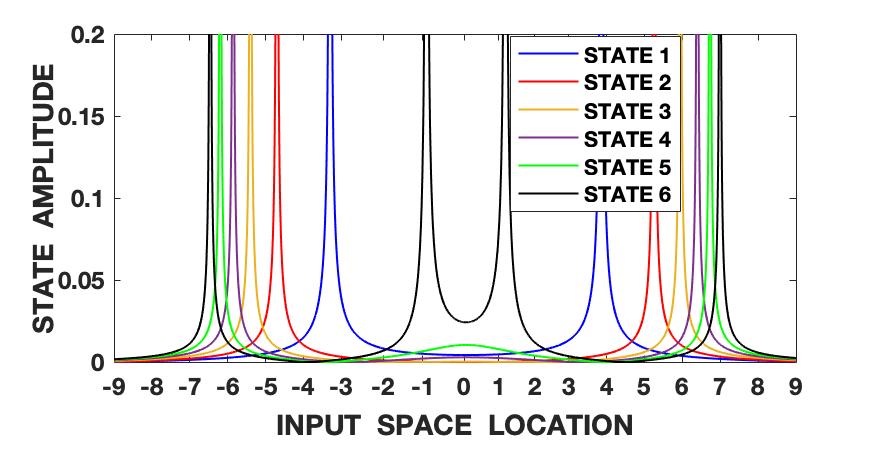}}
    \caption[caption]{Analysis of peak locations of the different QIPF modes in the input space at various kernel widths}
    \label{fig:5}
\end{figure*}

\setcounter{figure}{2}
\begin{figure*}[!b]
  \centering
    \subfigure[Signal (red) and points of evaluation for IPF and QIPF (blue)]{\includegraphics[scale = 0.1]{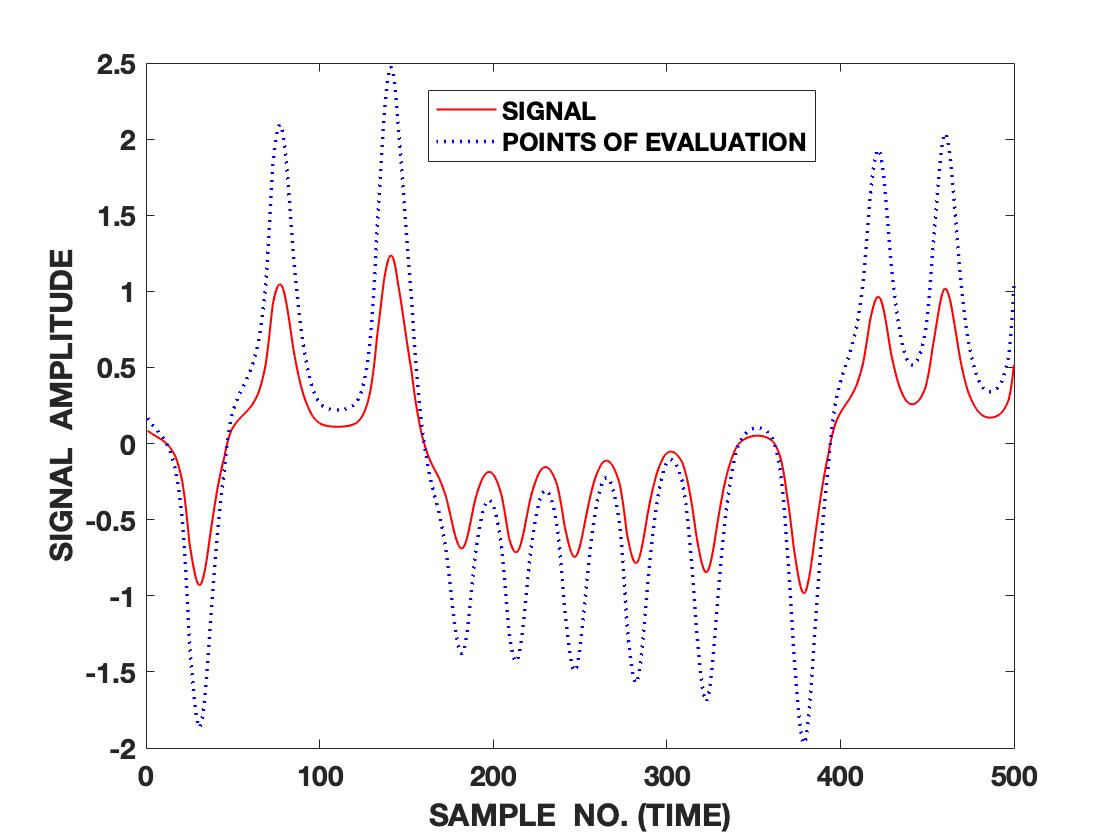}}
    \subfigure[Classical IPF and average of 5 QIPF modes]{\includegraphics[scale = 0.18]{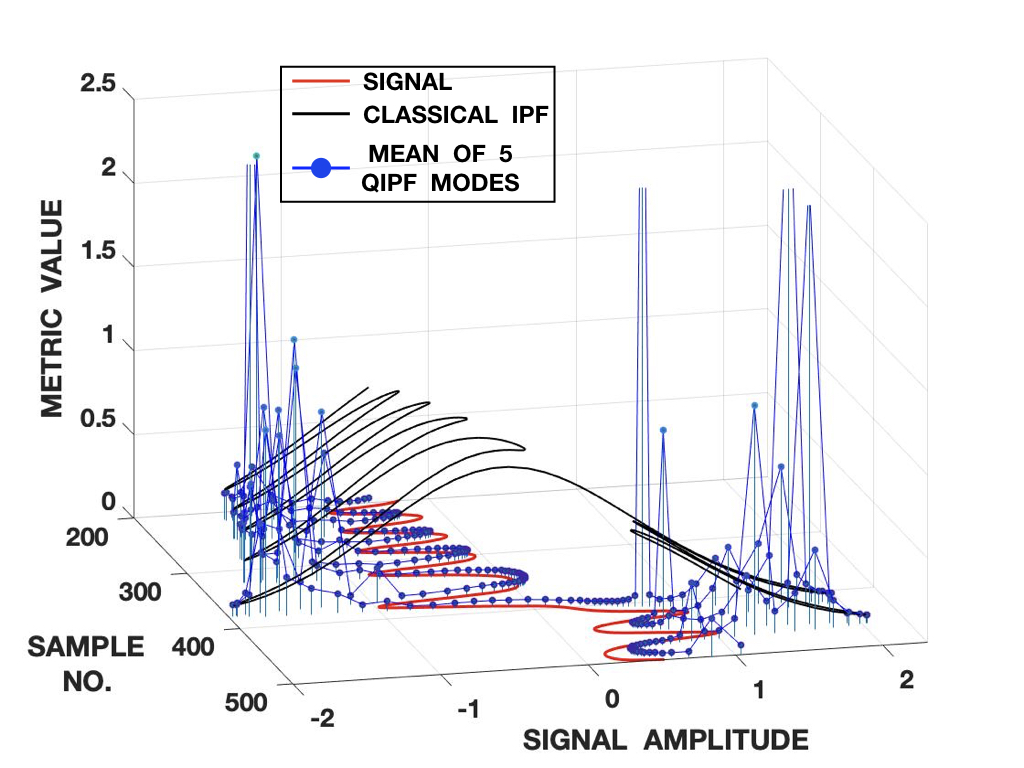}} 
    \subfigure[Classical IPF and average of 10 QIPF modes]{\includegraphics[scale = 0.18]{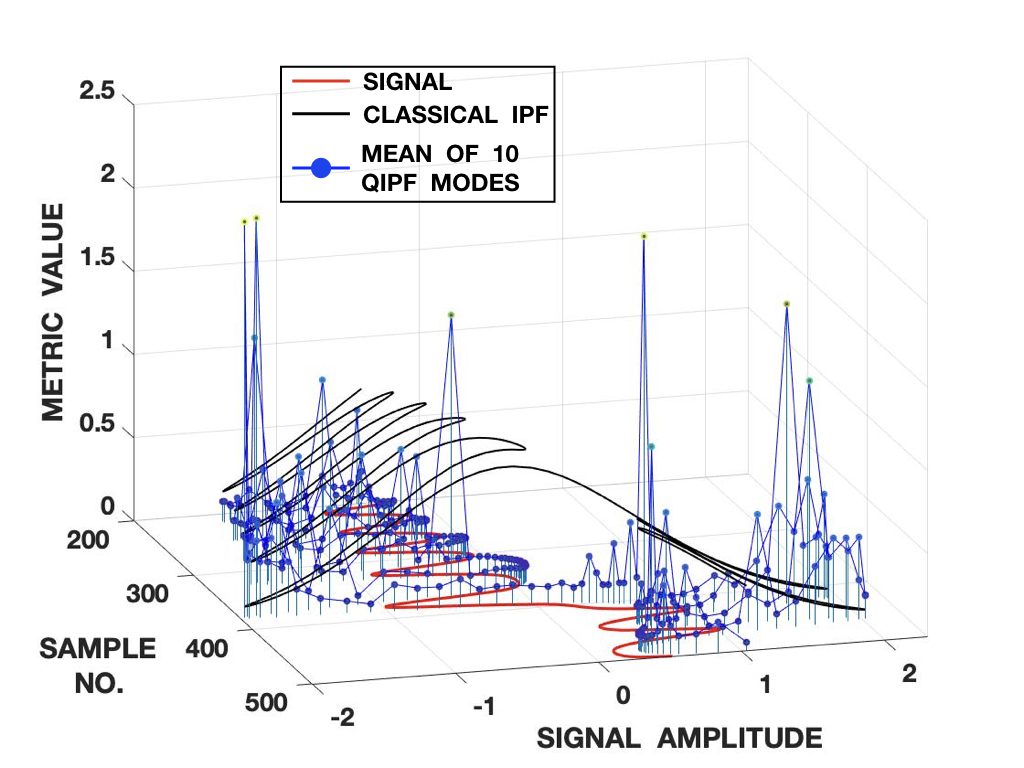}}
  \caption[caption]{Comparison of conventional information potential field and the average value of QIPF modes evaluated at different locations using past samples of a section of Lorenz series}
  \label{fig:1}
\end{figure*}

The Lorenz series is generated with the parameters set as $\sigma=10$, $\rho=28$ and $\beta=8/3$. The initial conditions are set as $x_1 = 0$, $y_1 = 1$ and $z_1 = 1.05$. The signal is also normalized to zero mean and unit variance and hence roughly varies between the amplitude levels of -2 and 2. We follow the steps shown in fig. 1 and algorithm 1 and first compute the fundamental wave-function (13) of the quantum information potential field (QIPF) at different locations in the input space using all of the 500 samples of the signal. This is followed by computation of the Hermite polynomial embeddings of the fundamental mode wave-function. We use 6 successive even order Hermite polynomial expansions for our analysis. We compute the QIPF corresponding to each of the extracted modes of the wave-function using (24) and (25), where $k$ represents the mode number and $i$ represents the sample number. Fig. 2 shows the plots of these modes extracted at a range of locations in the input space using different kernel widths. Several observations can be made from the plots. Firstly, one can observe that the different QIPF modes peak in a mutually exclusive manner and spread out upon increasing the kernel width. The mutual exclusivity can be attributed to the orthogonal nature of Hermite polynomial expansions. Secondly, we observe that the QIPF modes, in general, peak at their edges away from the mean and hence emphasize the more uncertain regions of the sample space. The higher order states successively emphasize the more distant regions in the input space. The trends of the different QIPF states in the region within the dynamic range of the signal are more sensitive to the kernel width. For extremely low kernel widths, as can be seen from fig. 2(a), the first few QIPF modes are completely diminished and the higher order modes dominate the signal's representation in all regions of sample space. Since the local kernel space of the signal modeled using extremely low kernel widths will be very uncertain, it is expected for higher order modes to dominate the signal's representation in such cases. On increasing the kernel width (fig. 2(b)), some lower order modes (modes 2 and 3) begin to dominate the signal's representation in the region around the mean. For moderate kernel widths (figs. 2(c) and 2(d)), the first mode of the QIPF dominates the immediate region around the mean. For larger kernel widths (figs. 2(e) and 2(f)) which exceed the dynamic range of the signal, we can see that high order modes begin to emerge in the region around the mean. Since very large kernel widths imply equally likely outcomes (increased uncertainty), it is logical for the higher order modes to dominate the signal's representation around the mean region. This behavior is remarkably similar to physical systems. Suppose that we have a drum, with a skin membrane. If we increase the tension of the membrane and hit it, the drum will vibrate for a long time. In our potential field, the stiffness is controlled by the kernel size. If the kernel size is large, the QIPF becomes stiffer leading to the energy in the higher QIPF modes to increase. If one decreases the kernel size, the membrane becomes more elastic leading to many local modes that decay much faster. One remarkable variable that we have not yet fully identified is the wavelength of the QIPF propagation, but its dependency on the kernel size is clear.\par

We analyze how the different regions and local dynamics of a signal are represented by our framework in real time by doing a sample-by-sample implementation (a causal analysis without access to the entire signal beforehand). As an illustrative example to this end, we implement our framework on 500 samples of Lorenz series generated and normalized in the same way as before. However, this time, the signal is scaled to half its original amplitude level. The samples of the scaled signal are used for computing the QIPF. The points at which QIPF is computed, however, are sample locations of the unscaled signal as is shown in fig. 3(a). This is done to evaluate the QIPF at regions outside of the dynamical range of the signal (in addition to the inside ones). It should be noted that the points of evaluation inside the dynamical range of the signal do not generally overlap with the scaled signal. They do, however, have roughly the same dynamical structure at the corresponding points. We evaluate the QIPF at each point of evaluation by extracting the QIPF modes and taking their average. We use a moderate kernel width of 0.7. At each point of evaluation, only the past samples of the scaled signal are utilized for the computations at that point. Fig. 3(b) shows the average of first 5 extracted modes of the QIPF at each point of evaluation along with the classical (conventional) information potential field values and fig. 3(c) shows the same for the first 10 modes of the QIPF. For visual clarity, only values from 200th to 500th points of\setcounter{figure}{3}
\begin{figure}[!t]
    \centering
    \subfigure[Sine Wave (100 Hz)]
    {\includegraphics[height = 1.5cm, width = 4.2cm]{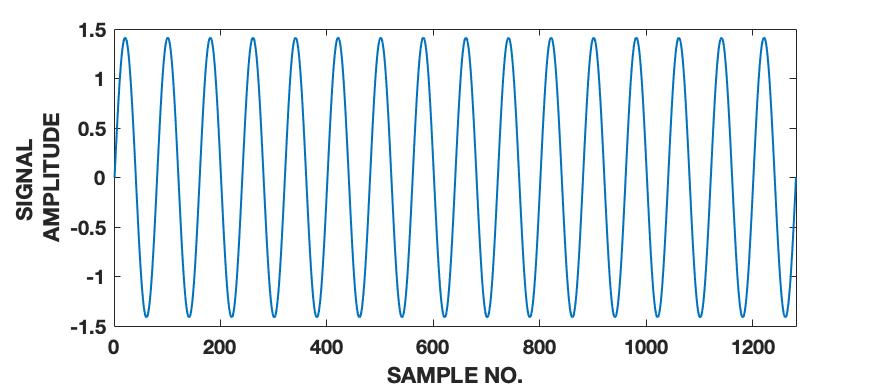}
    \includegraphics[height = 1.5cm, width = 4.2cm]{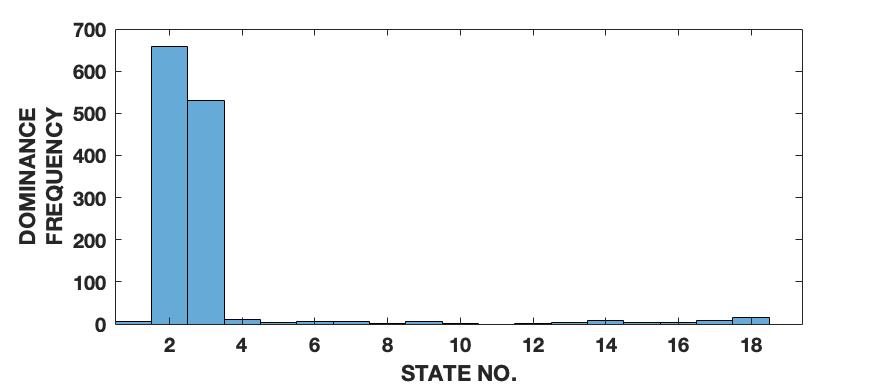}}
    
    \subfigure[Lorenz Series]
    {\includegraphics[height = 1.5cm, width = 4.2cm]{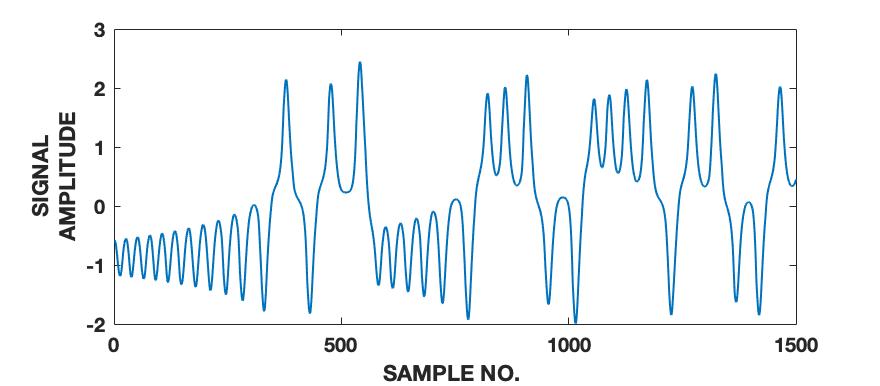}
    \includegraphics[height = 1.5cm, width = 4.2cm]{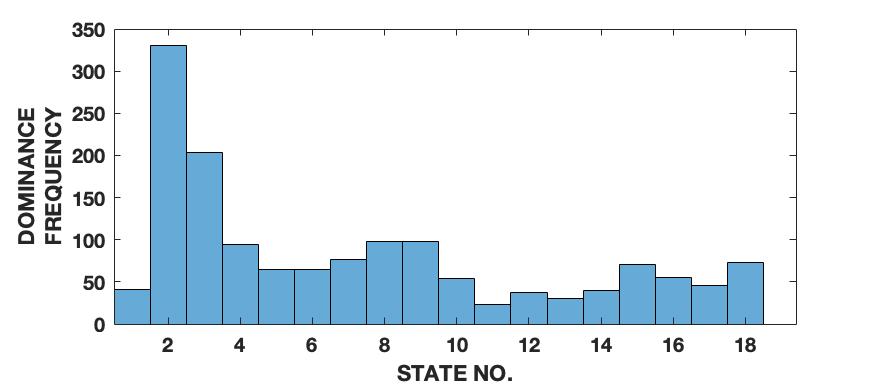}}
    \caption{Left: Generated signals. \hspace{\textwidth} Right: Corresponding dominance frequencies of QIPF states}
    \label{fig:2}
\end{figure}\setcounter{figure}{4}
\begin{figure}[!b]
\captionsetup[subfigure]{labelformat=empty}
    
    \centering
    \subfigure[$f_0 = 100$ Hz, $f_s = 8000$ Hz]
    {\includegraphics[height = 1.5cm, width = 4.2cm]{sine_100_sig.jpg}
    \includegraphics[height = 1.5cm, width = 4.2cm]{sine_100.jpg}}

    \centering
    \subfigure[$f_0 = 300$ Hz, $f_s = 8000$ Hz]
    {\includegraphics[height = 1.5cm, width = 4.2cm]{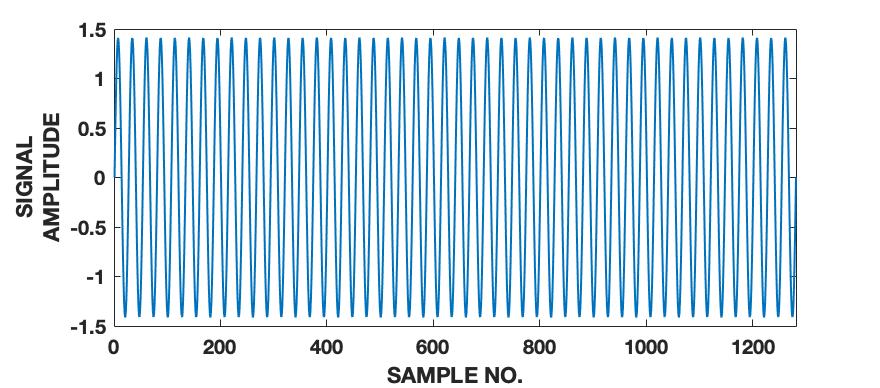}
    \includegraphics[height = 1.5cm, width = 4.2cm]{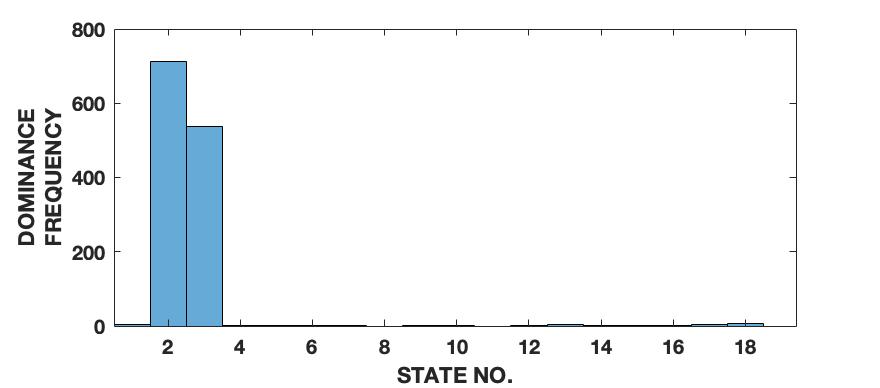}}
    
    \centering
    \subfigure[$f_0 = 300$ Hz, $f_s = 500$ Hz (Aliased)]
    {\includegraphics[height = 1.5cm, width = 4.2cm]{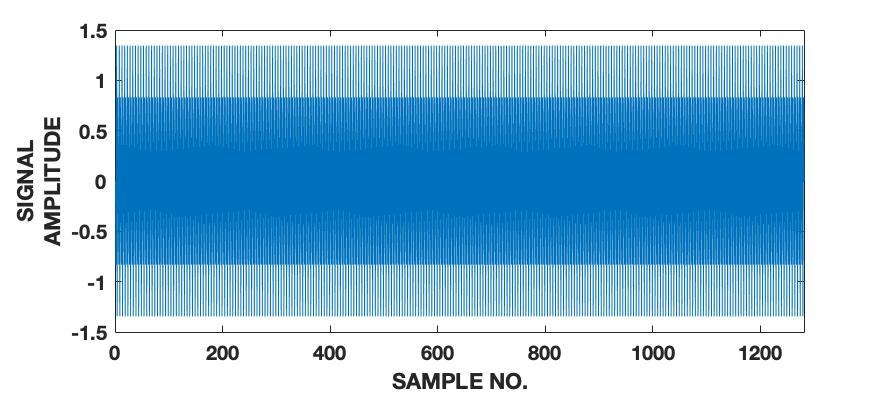}
    \includegraphics[height = 1.5cm, width = 4.2cm]{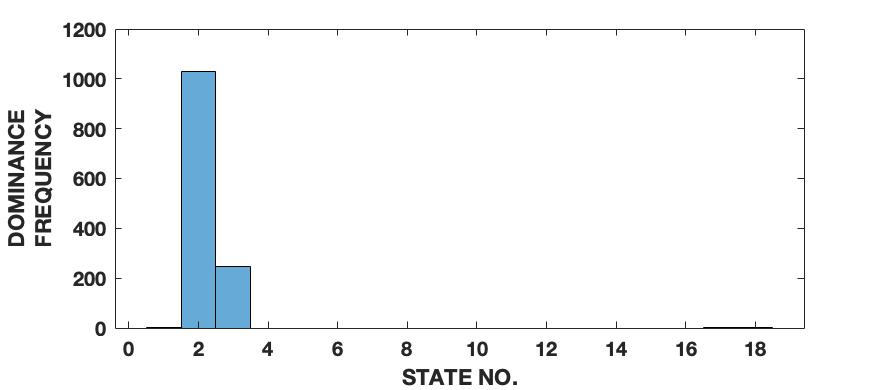}}
    
    \caption{Left: Different sine waves. Right: Corresponding dominance frequency of QIPF states.}
    \label{fig:3}
\end{figure}evaluation are shown for both cases. We can observe that in both cases, there is a significant difference between the sensitivity of the QIPF and that of the classical IPF. They also follow opposite trends with respect to the points of evaluation since the conventional (classical) IPF follows the PDF of the signal whereas the QIPF follows the uncertainty. The QIPF can also be seen to drastically increase at points of evaluation that are outside of the dynamical range of the scaled signal. It tends to be the lowest at high sample density regions. Upon comparing fig. 3(b) and fig. 3(c), one can notice that increasing the number of modes of the QIPF leads to a more detailed evaluation of how uniquely uncertainty gets quantified at every point (even within the dynamical signal range). One can observe more jumps/peaks in the uncertainty within the dynamical signal range when the number of modes is increased, especially at points where the signal dynamics visibly change.\par

\setcounter{figure}{5}
\begin{figure}[!b]
\captionsetup[subfigure]{labelformat=empty}
    
    \centering
    \subfigure[Freq. components: 300 Hz, 500 Hz]
    {\includegraphics[height = 1.5cm, width = 4.2cm]{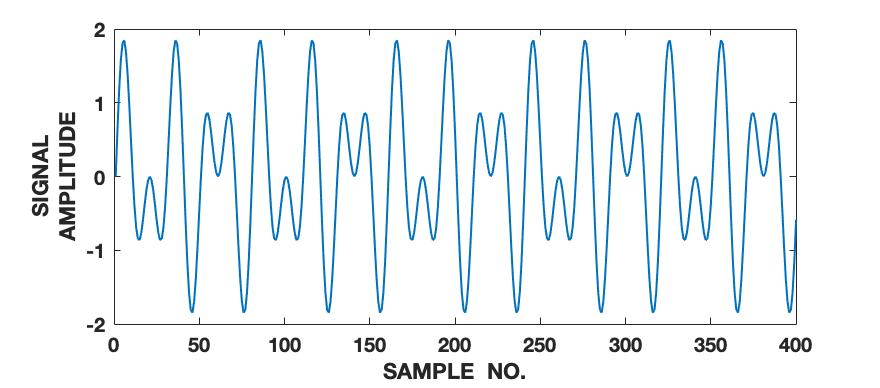}
    \includegraphics[height = 1.5cm, width = 4.2cm]{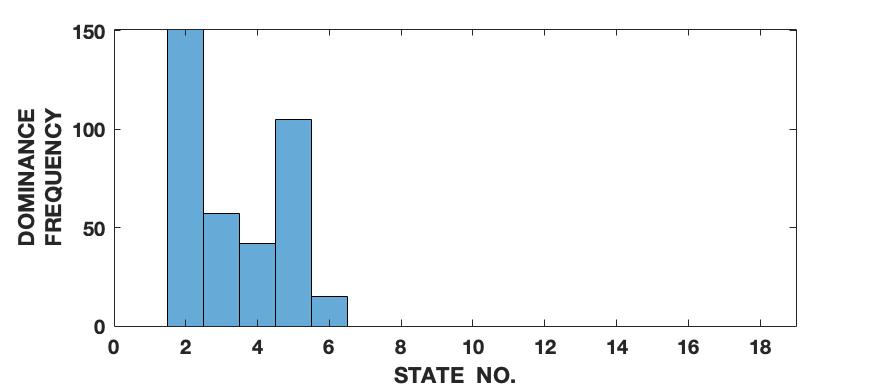}}
    
    \centering
    \subfigure[Freq. components: 100 Hz, 200 Hz, 300 Hz, 500 Hz, 750 Hz]
    {\includegraphics[height = 1.5cm, width = 4.2cm]{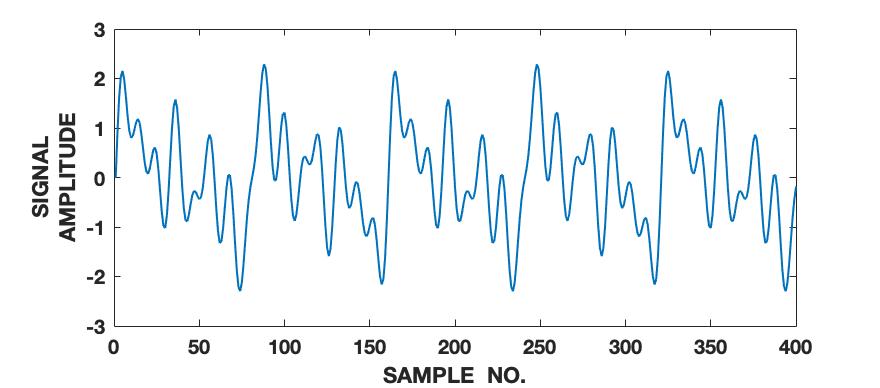}
    \includegraphics[height = 1.5cm, width = 4.2cm]{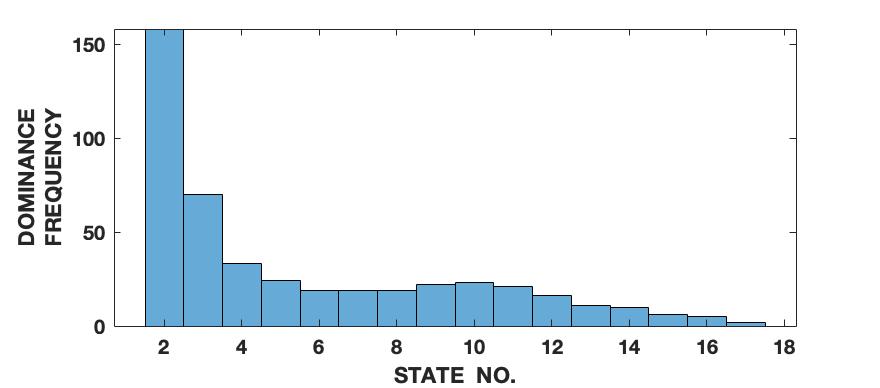}}
    
    \caption{Left: Sine waves with mixed frequency components. Right: Corresponding dominance frequency of QIPF states.}
    \label{fig:3}
\end{figure}

To demonstrate the effectiveness of the proposed framework in characterizing a signal in terms of its various intrinsic dynamical modes, we use a pedagogical example of how a sine wave function differs from Lorenz series with respect to their composition of dominant QIPF modes. Since our framework in based on the decomposition of local kernel space structure in terms of even order oscillator harmonics, we expect the sine wave, which consists of a single oscillator generating function, to get encoded in much fewer QIPF modes when compared to Lorenz series which is a chaotic dynamical system. We generate a sine wave having a frequency of 100 Hz and sampled at a rate of 8000 samples per second to mimic a continuous signal for a total time of 0.16 seconds. The Lorenz series is generated in the same way as before (without amplitude scaling). Both signals are normalized to zero mean and unit variance. We compute the QIPF modes using (24) and (25) and extract 18 successive even order modes of the QIPF to encode the uncertainties at the different sample locations of each signal. The kernel width used for doing so for both signals is fixed to 0.3, which is sufficiently small for more emphasis of the local regions of the sample space. Fig. 4 shows the signals (left column) and the corresponding histogram plots (right column) of the number of times the value of each QIPF mode dominated over the others throughout the durations of the signals. As can be seen in fig. 4, there are only two dominant modes in case of the sine wave (modes 2 and 3). The dominant modes of Lorenz series, on the other hand, are significantly more spread out towards higher orders thus indicating a more complex dynamical structure of the signal.\par

\setcounter{figure}{6}
\begin{figure}[!t]
    \centering
    \subfigure[($\sigma=10$, $\rho=28$, $\beta=8/3$); ($x_1 = 0$, $y_1 = 1$, $z_1 = 1.05$)]
    {\includegraphics[height = 1.5cm, width = 4.2cm]{lor_sig.jpg}
    \includegraphics[height = 1.5cm, width = 4.2cm]{lor.jpg}}
    \subfigure[($\sigma=10$, $\rho=28$, $\beta=5/3$); ($x_1 = 0$, $y_1 = 1$, $z_1 = 1.05$)]
    {\includegraphics[height = 1.5cm, width = 4.2cm]{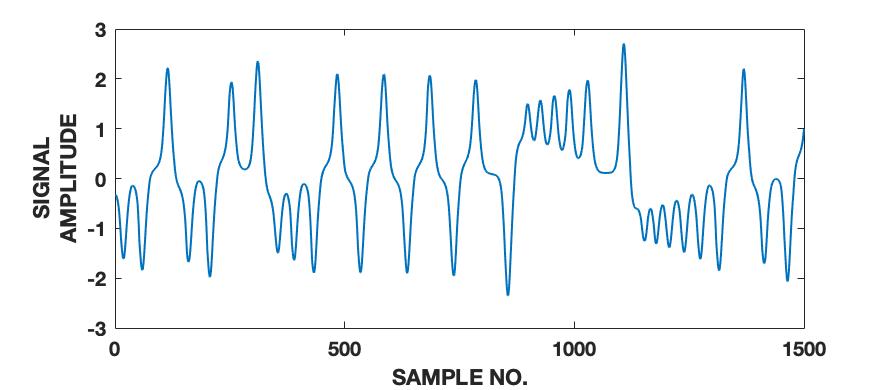}
    \includegraphics[height = 1.5cm, width = 4.2cm]{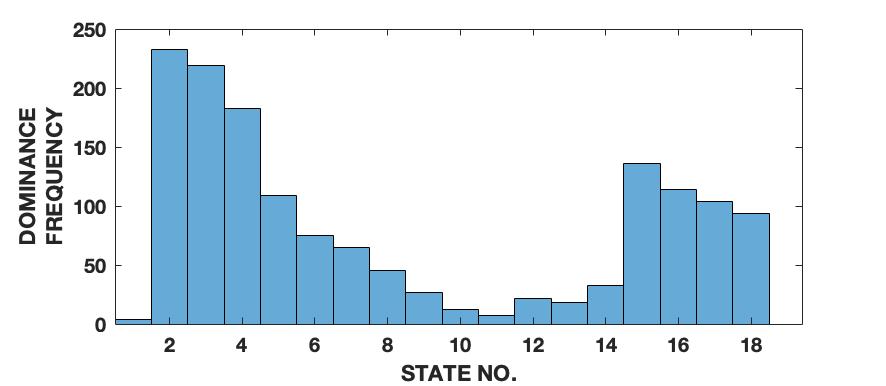}}
    \subfigure[($\sigma=10$, $\rho=28$, $\beta=8/3$); ($x_1 = 0$, $y_1 = 3$, $z_1 = 1.05$)]
    {\includegraphics[height = 1.5cm, width = 4.2cm]{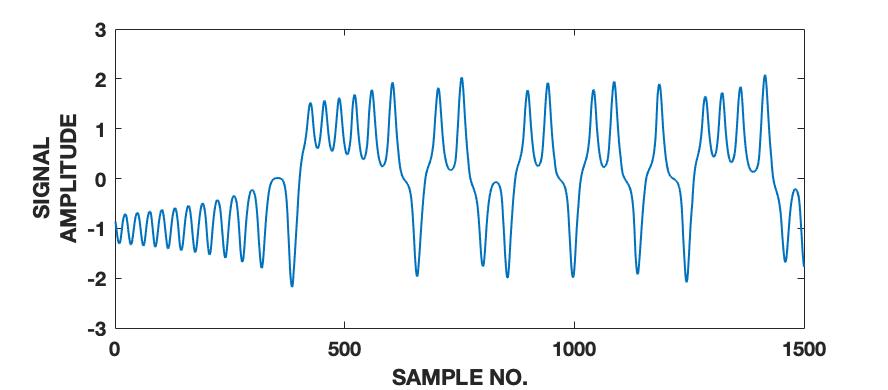}
    \includegraphics[height = 1.5cm, width = 4.2cm]{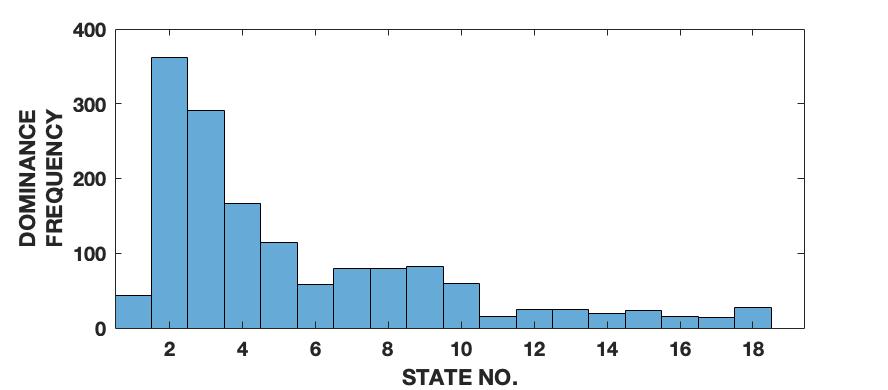}}
    \caption{Left: Lorenz series of different parameters or initial conditions. Right: Dominance frequency of QIPF states.}
    \label{fig:4}
\end{figure} 

We extend this analysis to different frequencies and sampling rates of the sine wave signal (figs. 5 and 6) as well as different parameters and initial conditions of the Lorenz series (fig. 7). It is interesting to note in fig. 5 that the effect of increasing the signal's frequency ($f_0$) has no significant effect on the distribution of QIPF modes that dominate the signal (which is still limited to modes 2 and 3). There is, however, a slight reduction in the third mode's proportion when the signal is aliased as can be seen in fig. 5(c). Fig. 6 shows the same analysis done on sine waves generated using a mixture of different frequency components. As can be seen from the corresponding histogram plots, the distribution of dominant QIPF modes begins to spread out more towards the higher modes when the number of frequency components in the signal increases. The modal distribution of the Lorenz series, in fig. 7, can be seen to be significantly more responsive towards changing generating system parameters and initial conditions. Since chaotic dynamical systems have a high sensitivity and long-term dependency towards system parameters and initial conditions, this result is expected. Overall, these results show that our framework is more dependent on the genrating system parameters than local changes when charaterizing signals in terms of composition of modes.\par

\begin{table}[!b]
  \begin{center}
    \caption{Noise Parameters}
    \label{tab:table1}
    \begin{tabular}{c c c} 
    \toprule
      \textbf{Sample Interval} & \textbf{SNR (DB)} & \textbf{Variance}\\[0.7ex] 
      \midrule
      500-600 & 16.7 & 0.021\\
      \addlinespace
      
      600-700 & 20.4 & 0.009\\
      \addlinespace
      700-800 & 14.2 & 0.038\\
      \addlinespace
      800-900 & 16.3 & 0.034\\
      \addlinespace
      900-1000 & 14.5 & 0.035\\
      \addlinespace
      1000-1100 & 5.5 & 0.281\\
      \addlinespace
      1100-1200 & 10.3 & 0.093\\
      \bottomrule
    \end{tabular}
  \end{center}
\end{table}

We also compute the QIPF energy levels (or eigenvalues) of the first 18 even order modes of the information potential extracted from various signals. The formulation for computing the energy levels associated with various states is given by (25), which is the minimum value of the Laplacian at each mode required to constraint the QIPF at that mode to be positive throughout time. Fig. 8(a) shows the normalized energy levels associated with the first 18 even order states of sine wave signals of different frequencies. All of these signals are generated using the same sampling rate (8000 Hz) and the samples in the first 0.05 seconds of each signal are considered for the computations. The eigenvalues shown here for each mode are computed using all of the samples simultaneously. Fig 8(b) similarly shows the normalized QIPF energy levels associated with the first 18 even order QIPF modes of the Lorenz series signals generated using different parameters or initial conditions. 500 samples for each Lorenz series signal are used for the evaluations. It can be seen from both subfigures in fig. 8 that the energy levels increase consistently with the state number. It is interesting to note here that the slope of the energy level vs state number curve generally flattens out rather quickly (within first 4-5 modes) for the sine wave signals as compared to that associated with the Lorenz series signals. This suggests (and is also expected) that higher order states do not contribute significantly in the representation of the sine wave dynamics as they do in the representation of Lorenz series signals. This is also consistent with the trends observed in fig. 5. For increased frequency sine wave, however, larger variance is observed in the energy levels of higher order states, likely due to increasingly varying characteristics and sensitivities of the higher order modes. For both the classes of signals in fig. 8, one can observe that the first few modes of the QIPF are quite similar. Discriminative properties particular to specific signals (caused due to change in parameters, initial conditions or fundamental frequency components) start to get reflected in higher order QIPF states.\par

\setcounter{figure}{7}
\begin{figure}[!t]
  \centering
    \subfigure[Sine Wave]{\includegraphics[height = 4.1 cm, width=0.49\linewidth]{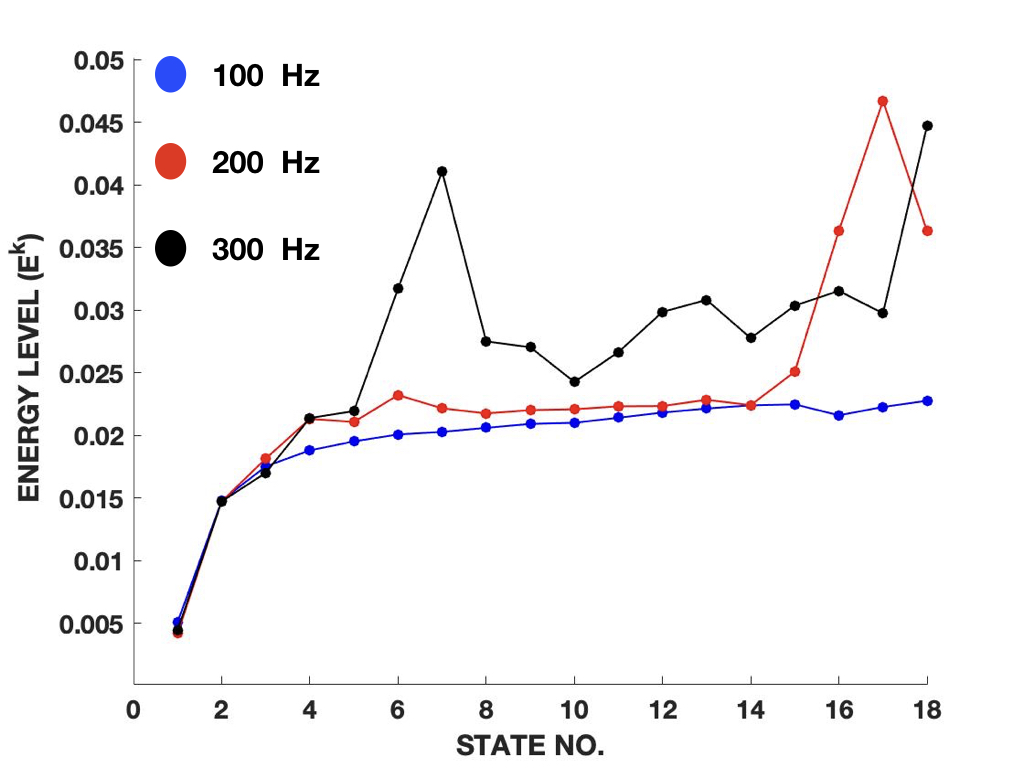}}
    \subfigure[Lorenz Series]{\includegraphics[height = 4.1 cm, width=0.49\linewidth]{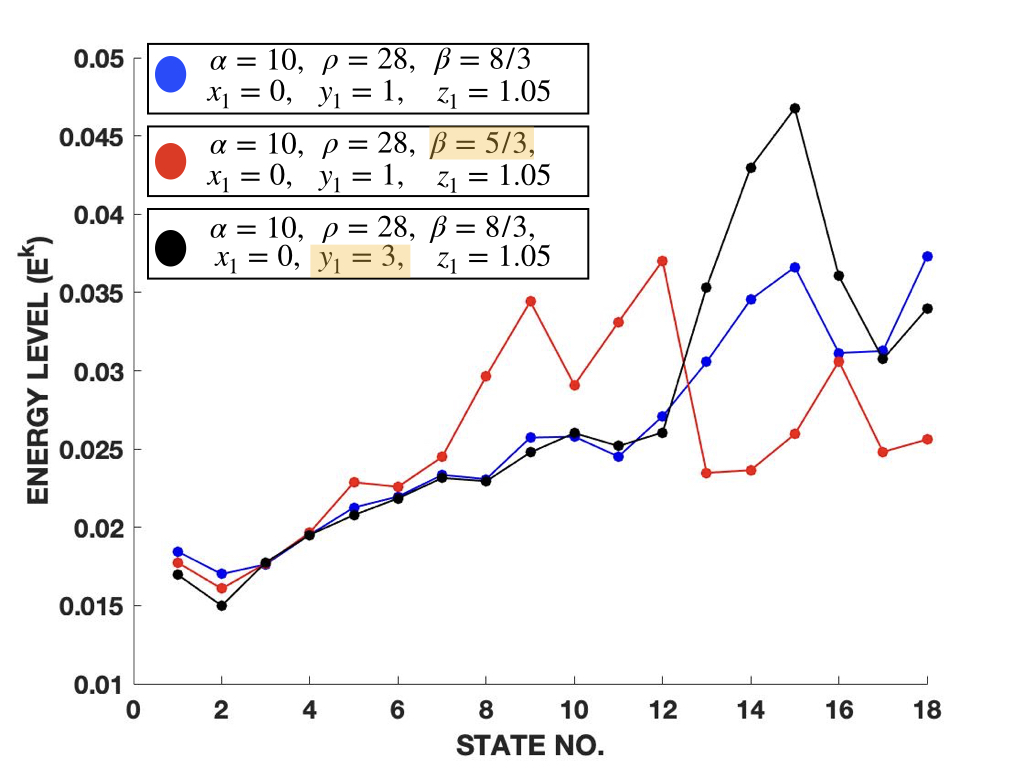}}
  \caption[caption]{Normalized eigenvalues at different QIPF states}
  \label{fig:6}
\end{figure}

\setcounter{figure}{8}
\begin{figure}[!b]
  \centering
\includegraphics[height = 1.9cm, width=7cm]{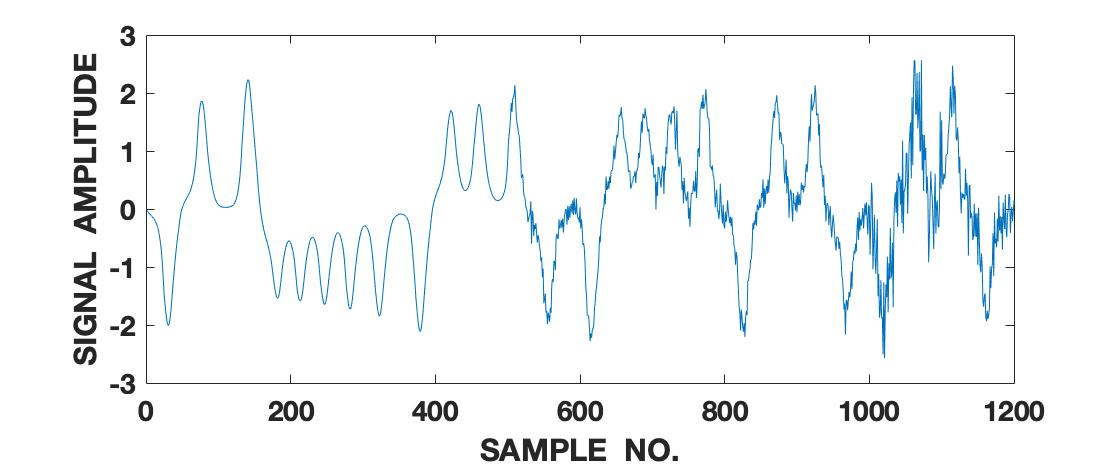}      
\caption{Addition of heteroscedastic noise in Lorenz series}
  \label{fig:7}
\end{figure}
 
\setcounter{figure}{9}
\begin{figure*}[!t]
\centering
\includegraphics[width= 17 cm, height = 6 cm]{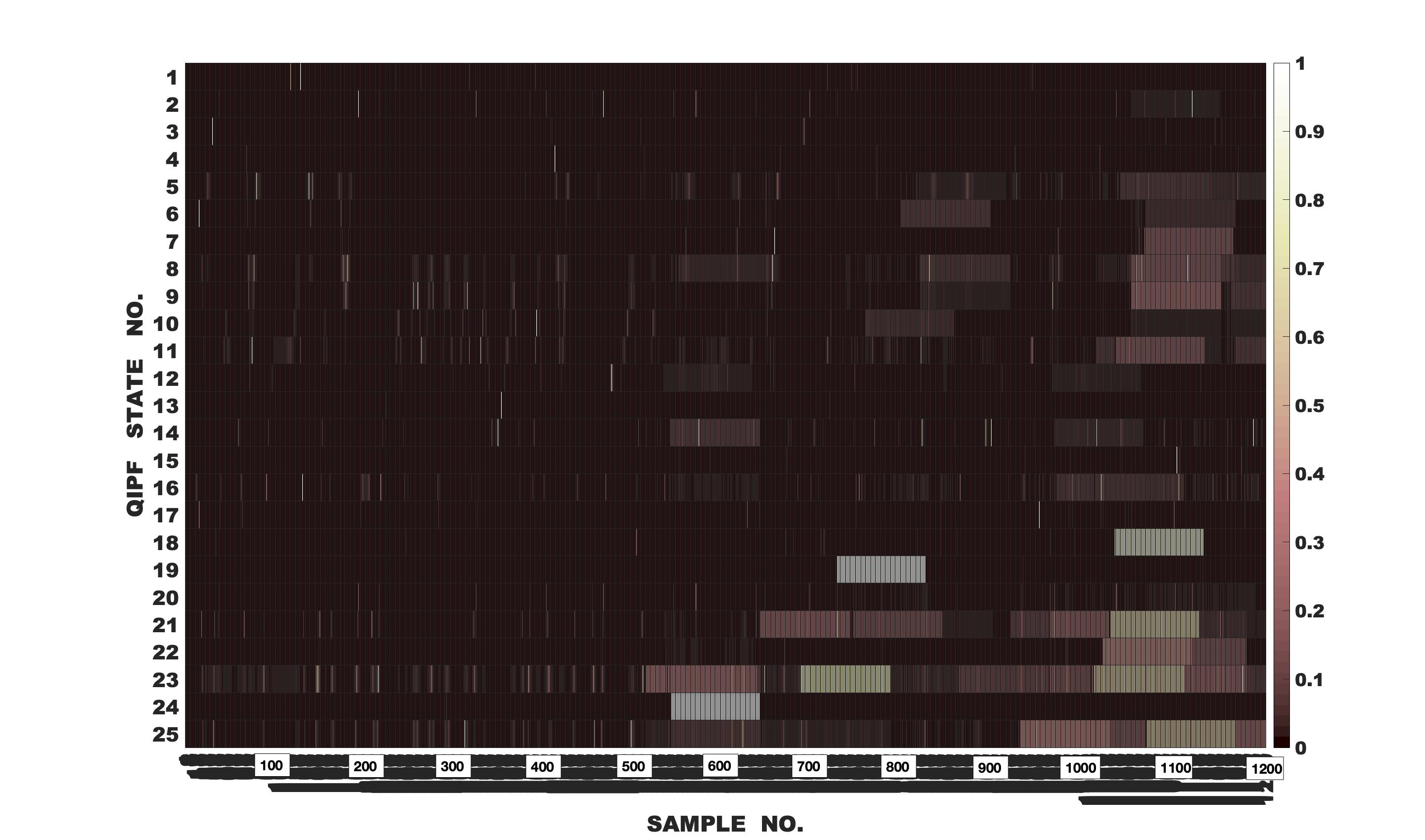}
\caption{Normalized heat-map depicting values of different QIPF states at each sample}
\label{fig:8}
\end{figure*}

\setcounter{figure}{10}
\begin{figure}[!b]
  \centering
  \includegraphics[height = 6.5 cm, width=8 cm]{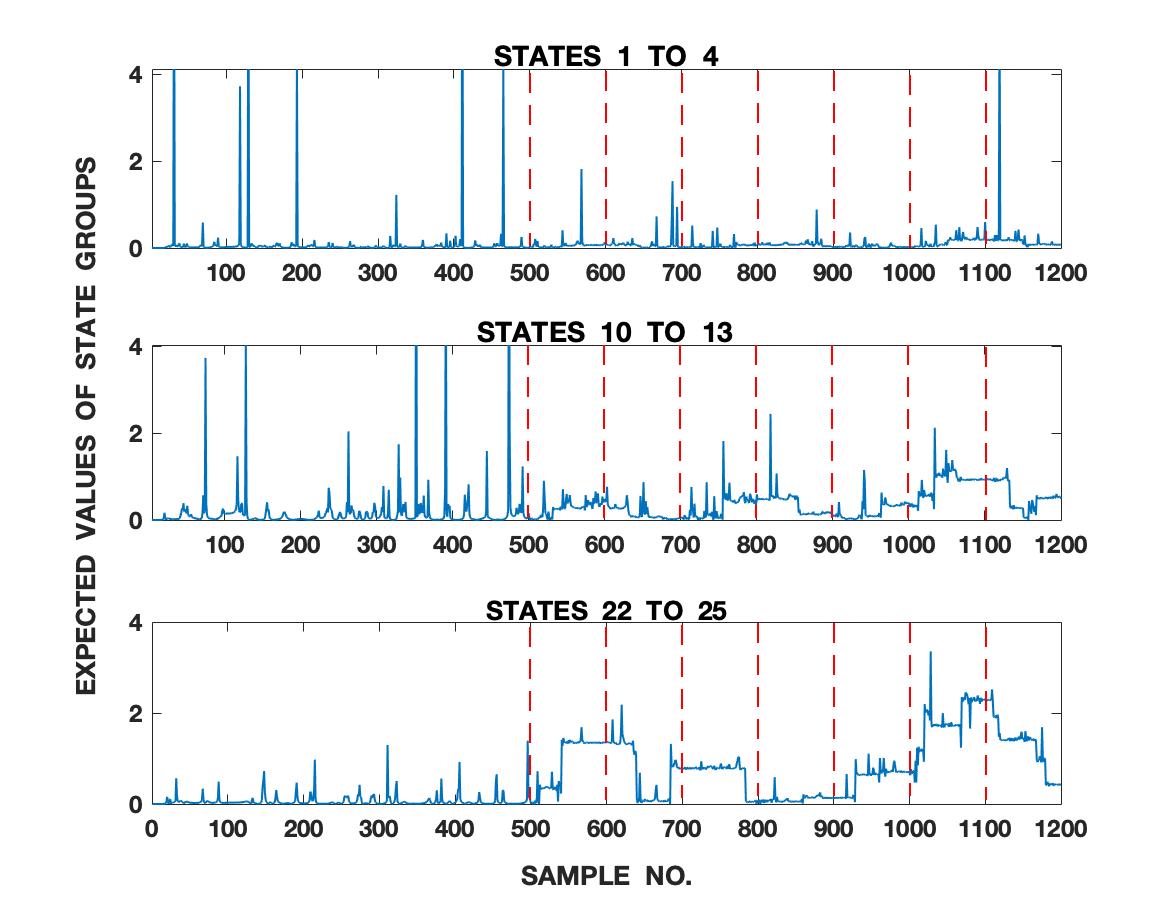}
  \caption{Average values of different groups of QIPF states (dotted red lines indicate noise intervals)}
  \label{fig:9}
\end{figure}

In order to illustrate the sensitivity of the proposed framework on the changing dynamical behavior and uncertainties associated with the signal, we implement our framework on a sample-by-sample basis on 1200 samples of Lorenz series where the last 700 samples are corrupted using heteroscedastic white Gaussian noise. The noise variance (and consequently the SNR) randomly changes after every 100 samples (as shown in table 1). The first 500 samples are uncorrupted. We use a window of past 100 samples to perform computations at each sample location. The signal under consideration is shown in fig. 9. The noise is added from the 500th sample onwards. We compute the first 25 states of the QIPF at every sample location using the same formulations given by (24) and (25) and as depicted in fig. 1. We use a kernel width of 0.4. The heat-map shown in fig. 10 represents the values of the different states of the QIPF at all sample locations. For clarity of analysis, all values in the heat-map matrix are normalized with respect to values only in its corresponding row (i.e. all values of the particular state throughout time). From the heat-map, we can observe how the different states of the QIPF react to the inclusion of noise. The first few states (1 to 5) show very little response to noise changes in general. As we move on to the middle order states (6 to 15), we observe increased response to some high changes in the noise statistics. Higher order states (state 16 onwards) can be seen to be even more sensitive towards changes in noise statistics. The highest order states can be seen to respond drastically towards all types of noise inclusion from the very first interval itself. Fig. 11 shows the expected value of the different sets of QIPF states (1-4, 10-13 and 22-25) at each sample location. The sensitivity trends of the different sets of states are apparent here with the expected value of the first 4 states changing negligibly with noise inclusion. The expected value of the middle order states (10 to 13) can be seen to be significantly more responsive towards some of the high noise transitions (especially from the 1000th sample onwards). Expected value of the higher order states (22 to 25) shows increased sensitivity towards all changes in noise variances as is evident from its relatively quick response towards all intervals of noise. Furthermore, the changes in the expected value of higher order states can also be seen to be roughly correlated with the changes in SNR values asscociated with the noise. It is interesting to observe here that addition of noise does not lead to more oscillations or spikes in the expected state values even though the framework is implemented on a sample-by-sample basis. At the same time, however, the expected value of higher order states changes significantly at different noise intervals. This is indicative of the framework's ability to identify the global signal statistics despite operating locally.\par
 
 \begin{table*}[!t]
\begin{center}
    \begin{tabular}{c c c c c c c c c c c c c c c c}
    \toprule
    \addlinespace
    \multicolumn{5}{c}{} & \multicolumn{11}{c}{\textbf{Kernel Width}} \\
    \addlinespace
    \multicolumn{2}{c}{\textbf{Framework}} & & & & \textbf{0.2} & & \textbf{0.4} & & \textbf{0.5} & & \textbf{0.6} & & \textbf{0.8} & & \textbf{1} \\
    \addlinespace
    \midrule
    \addlinespace
    & States 1-3 & & & & 1.1860  & & 1.0996 & & 0.9047 & & 0.5578 & & \textbf{2.5123}  & & \textbf{1.7937}\\ 
    \addlinespace
   QIPF & States 4-6 & & & & 1.4799 & & 0.8292 & & 2.8727 & & \textbf{4.4312} & & 1.7927  & & 1.6161\\ 
   \addlinespace
    & States 7-10 & & & & 1.4956 & & \textbf{3.7847} & & \textbf{3.7879} & & 1.4028 & & 1.2349 & & 1.2133\\
    \addlinespace
    \midrule
    \addlinespace
    \multicolumn{2}{c}{Bayesian Surprise} & & & & 1.8301 & & 1.6548 & & 1.6681 & & 1.6255 & & 1.2913 & & 0.5803\\
    \addlinespace
    \midrule
    \addlinespace
    \multicolumn{2}{c}{Entropy Difference} & & & & \textbf{1.9174} & & 1.7482 & & 1.5896 & & 1.4160 & & 1.0191 & & 0.7853\\
    \addlinespace
    \midrule
    \addlinespace
    \multicolumn{2}{c}{Classical IP} & & & & 0.3174 & & 0.2942 & & 0.2644 & & 0.2451 & & 0.2343 & & 0.2400\\
    \addlinespace
    \bottomrule

    \end{tabular}
    \caption{Sensitivity of several frameworks to changing noise statistics at different kernel widths}
    \end{center}
    
 \end{table*}

We use this approach to quantify the sensitivity of our framework towards statistical changes in the signal and thereby compare our framework with entropy based surprise quantification methods \cite{17}, \cite{18}. We perform the quantification and analysis using our framework on 5000 samples of Mackey Glass chaotic dynamical series. This dynamical system is governed by the following non-linear delay differential equation:

\begin{equation}
\frac{dx}{dt} = \alpha\frac{x(t-\tau)}{1+x(t-\tau)^n} - \beta{x} \; \; \; \; \; \; \alpha, \beta\ {and} \; n > 0
\end{equation}
where $\tau$ is the delay and $\alpha$, $n$ and $\beta$ are other parameters.\par

The 5000 samples are generated by setting the delay parameter ($\tau$) as 30 and other parameters as $\alpha = 0.2$, $\beta = 0.1$. Heteroscedastic noise (in the range of 0-20 decibels), with the variance changing after every 500 samples, is added to the last 2500 samples of the generated Mackey Glass signal. We implement our framework on the signal by extracting the first 10 even order QIPF modes followed by finding the expected values of states 1 to 3, 4 to 6 and 7 to 9 at each sample location denoted by $V_i^{1-3}$,$V_i^{4-6}$ and $V_i^{7-9}$ respectively, where $i$ is the sample number. We then quantify the change detection performance of the three groups of states by measuring their sensitivity with respect to the change in noise variance. The sensitivity ($\zeta$) is measured by evaluating the change in Euclidean norm of the state values from one interval of the noise corrupted samples to the next where the variance of the noise changes. It is given as:

\begin{equation}
\fontsize{8pt}{12pt}\selectfont
\zeta = E_i\bigg(\bigg |\frac{||{V_R^K}|| - ||{V^K_{R+1}}||}{D_R - D_{R+1}}\bigg|\bigg)
\end{equation}

Here $R$ is the sample interval and $K$ is 1-3, 4-6 or 7-9 depending on which state group we consider. $D_R$ is the decibel measure of noise in the sample interval $R$. Hence, using this measure, we evaluate the change in a particular state group with respect to the change in noise from one sample interval, $R$, to the next, $R+1$. The sample intervals are non-overlapping and have a length of 500 samples which is the same as the interval length in the heteroscedastic noise where the variance is constant. We compare the sensitivity of our framework with that of the Bayesian surprise model (7) which measures the amount of surprise associated with each data sample by evaluating the KL divergence between the prior and posterior model distributions. It is reformulated here as below:

\begin{equation}
\fontsize{8pt}{12pt}\selectfont
\centering
S(D, \mathscr{M}) =
\int\limits_{\mathscr{M}}P(M|D)log{\frac{1}{P(M)}}dM - \int\limits_{\mathscr{M}}P(M|D)log{\frac{1}{P(M|D)}}dM
\end{equation}

For comparisons, we use the Parzen density (using Gaussian kernel windows) as the model distribution $P(M)$ in the Bayesian surprise framework since the information potential, which is used by our framework as the local kernel space quantifier, is ultimately a metric derived from Parzen density estimation (as has been shown in section II). The model distribution in the Bayesian surprise framework is updated with every sample. However, the space over which integral computations are done at every sample (model space $\mathscr{M}$ in (29)) is fixed to be a regularly spaced set of values lying in the dynamic range of the generated Mackey Glass series. The sensitivity here is measured by replacing $V_R^K$ and $V_{R+1}^K$ in (28) with the two integral computations seen in (29). Here $P(M|D)$ is the set of updated distributions with respect to the data in the next interval of samples. Hence, sensitivity related to KL divergence evaluated here measures the differences in statistics of the neighboring interval of samples where the noise variances are different from each other. The other metric we compare our framework with is the sensitivity associated with simply the entropy differences between the different sample intervals where the noise variance changes. The sensitivity values for the  different frameworks evaluated at various kernel widths are shown in table II. The values shown are the average results of 10 simulation runs for each framework. Before evaluating the sensitivities, normalization of values (to zero mean and unit variance) was done for each framework with respect to the evaluated values at all samples and independent of other frameworks. It can be seen here that for all kernel widths above 0.2, our framework has higher sensitivity to changes in signal statistics than other models. Upon analyzing how the sensitivity of our framework changes with respect to kernel width, one can notice that the higher order set of QIPF states are more sensitive at lower kernel widths. On increasing the kernel width, the lower order QIPF states start becoming more sensitive when compared to higher order states. This is expected because increasing the kernel width starts to spread out the higher order states to beyond the dynamical range of the signal within which changes take place. However, there is also a compromise here because the best sensitivity values (among the different state groups) can be seen to decrease when the kernel width is increased thus indicating that the higher order states capture the changes in signal statistics better than the lower order states. Our framework is the most sensitive at moderate kernel widths of 0.5 and 0.6 (at which the higher order state group is most sensitive) for this example. In general, the variation of state sensitivity with respect to the kernel width also depends on the dynamcal structure of the signal.

\section{Conclusion}
In this paper, we have introduced a new framework for stochastic signal processing that utilizes the information potential as the local kernel space quantifier and exploits its quantum physical description to extract its various dynamical modes that quantify uncertainty. We have stressed on the importance of local metric space structures in the charaterization of signals, instead of probabilistic measures. We have shown how this framework can be implemented on a sample-by-sample basis and have highlighted its several key advantages compared to other methods both theoretically and experimentally. We have also shown that our framework is highly sensitive towards the local intrinsic dynamics of the signal while also being able to provide a generalized global stochastic representation. As future work, we intend to apply this framework in the context of adaptive filtering and predictive algorithms. We also intend to utilize this framework as an RKHS based uncertainty quantifier to analyze the performance and robustness of deep learning models.



\ifCLASSOPTIONcaptionsoff
  \newpage
\fi



\end{document}